\RequirePackage{fix-cm}
\documentclass[smallextended]{svjour3wide} 
\smartqed  
\usepackage{graphicx}
\usepackage{amsmath}
\usepackage{amssymb}
\usepackage{mathrsfs}
\usepackage{color}
\usepackage{bm}
\newcommand{\ave}[1]{\ensuremath{\left\langle#1\right\rangle}}
\newcommand{\aves}[1]{\ensuremath{\langle#1\rangle}}

\newcommand{\rd}{\partial}
\journalname{Journal of Statistical Physics}
\begin{document}

\title{Derivation of Stokes' Law from Kirkwood's Formula and \\ the Green--Kubo Formula via Large Deviation Theory}
\titlerunning{Derivation of Stokes' Law} 

\author{Masato Itami \and Shin-ichi Sasa}

\institute{M. Itami \and S. Sasa \at
              Department of Physics, Kyoto University, Kyoto 606-8502, Japan \\
              \email{itami@scphys.kyoto-u.ac.jp, sasa@scphys.kyoto-u.ac.jp}
}

\date{Received: date / Accepted: date}

\maketitle

\begin{abstract}
We study the friction coefficient of a macroscopic sphere in a viscous fluid at low Reynolds number.
First, Kirkwood's formula for the friction coefficient is reviewed on the basis of the Hamiltonian description of particle systems.
According to this formula, the friction coefficient is expressed in terms of the stress correlation on the surface of the macroscopic sphere.
Then, with the aid of large deviation theory, we relate the surface stress correlation to the stress correlation in the bulk of the fluid, where the latter is characterized by the viscosity in the Green--Kubo formula.
By combining Kirkwood's formula and the Green--Kubo formula in large deviation theory, we derive Stokes' law without explicitly employing the hydrodynamic equations.
\end{abstract}

\section{Introduction}

In equilibrium systems, macroscopic behavior is systematically described by a universal framework---thermodynamics~\cite{Callen}.
On the basis of a microscopic description of equilibrium systems, the principle of equal a priori probabilities and Boltzmann's principle reproduce the framework of thermodynamics, which is established as equilibrium statistical mechanics~\cite{Landau-LifshitzStat}.
In contrast, there is still no theory describing the general behavior of non-equilibrium systems beyond the linear response regime.
Thus, much effort has been devoted to the investigation of steady-state thermodynamics and non-equilibrium statistical mechanics~\cite{Groot-Mazur,Zubarev,Kubo-Toda-Hashitsume,Sasa-Tasaki}.

When liquids and gases are out of equilibrium but still remain in local equilibrium, their macroscopic dynamical behavior is precisely described by the hydrodynamic equations~\cite{Landau-LifshitzFluid}.
A microscopic understanding of the hydrodynamic equations for the case of dilute gases was established through the Boltzmann equation~\cite{Chapman-Cowling}, whereas it remains unclear for a general fluid.
Non-trivial relations that are generally valid far from equilibrium, including the fluctuation theorems~\cite{Evans-Cohen-Morriss,Gallavotti-Cohen,Kurchan,Lebowitz-Spohn,Maes,CrooksPRE1,CrooksPRE2,JarzynskiJSP} and the Jarzynski equality~\cite{JarzynskiPRL}, have been developed over the past two decades as a result of the time-reversal symmetry of microscopic mechanics.
Thanks to such universal relations, we can easily re-derive certain well-known relations, such as the McLennan ensembles, the Green--Kubo formula, and the Kawasaki nonlinear response relation~\cite{Hayashi-Sasa,SeifertRPP}.
Furthermore, by using a non-equilibrium identity similar to the fluctuation theorems and assuming a local Gibbs distribution at the initial time, the Navier--Stokes equation was derived for an isolated Hamiltonian system~\cite{Sasa}.
Based on these achievements, we believe this is an opportune moment to reconsider fluid dynamics from the viewpoint of statistical mechanics. 

In this paper, we study the friction coefficient of a macroscopic sphere in a fluid.
Starting from microscopic mechanics, Kirkwood first derived the linear response formula for the friction coefficient, $\gamma_{\rm K}$, in the form~\cite{Kirkwood}
\begin{equation}
 \gamma_{\rm K} = \frac{1}{3k_{\rm B}T}\int_{0}^{\tau}dt\; \ave{\bm{F}_{t}\cdot \bm{F}_{0}}_{\rm eq},
  \label{eq:kirkwood_formula_intro}
\end{equation}
where $k_{\rm B}$ is the Boltzmann constant, $T$ is the temperature of the fluid, $\aves{\; \cdot \;}_{\rm eq}$ denotes a canonical ensemble average at temperature $T$, and $\bm{F}_t$ is the total force exerted on the sphere by the fluid at time $t$.
The upper limit of the integral $\tau$ should be much larger than the correlation time of the force $\bm{F}_{t}$ and much smaller than the relaxation time of the momentum of the sphere.
This will be discussed in detail in Sect.~\ref{sec:kirkwood}.
Note that $\tau$ could go to infinity if we take the heavy mass limit for the sphere~\cite{Lebowitz-Rubin,ZwanzigJCP,Mazur-Oppenheim}.
The theoretical results were confirmed by numerical experiments~\cite{Lagarkov-Sergeev,Brey-Ordonez,Espanol-Zuniga,Kaddour-Levesque,Lee-Kapral}.

By focusing on a special class of fluids, we can obtain a more explicit expression for the friction coefficient.
For instance, the friction coefficient for a dilute gas, $\gamma_{\rm dg}$, has the form~\cite{Lorentz,Green1951}
\begin{equation}
 \gamma_{\rm dg}  = \frac{8}{3}\rho \mathcal{R}^2\sqrt{2\pi m k_{\rm B}T},
  \label{eq:kinetic_fric_intro}
\end{equation}
where $\rho$ is the number density of gas particles, $\mathcal{R}$ is the radius of the sphere, and $m$ is the mass of a gas particle.
As another typical example, the friction coefficient for a viscous fluid at low Reynolds number, $\gamma_{\rm S}$, is given by~\cite{Landau-LifshitzFluid}
\begin{equation}
 \gamma_{\rm S} = \mathcal{C} \eta \mathcal{R},
  \label{eq:stokes_fric_intro}
\end{equation}
where $\eta$ is the viscosity of the fluid.
The numerical coefficient $\mathcal{C}$ is equal to $4\pi$ for the slip boundary condition where the velocity of the fluid normal to the sphere at the boundary is equal to that of the sphere in this direction and the shear stresses on the sphere are equal to zero, and $6\pi$ for the stick boundary condition where the velocity of the fluid at the boundary is equal to that of the sphere.
The expression (\ref{eq:stokes_fric_intro}) is known as Stokes' law.

Kirkwood's formula can be derived on the basis of a mechanical system, meaning that the validity of the formula is generally independent of the nature of the fluid within the linear response regime.
Thus, Kirkwood's formula (\ref{eq:kirkwood_formula_intro}) corresponds to (\ref{eq:kinetic_fric_intro}) for a dilute gas, which was confirmed by Green~\cite{Green1951}.
Furthermore, (\ref{eq:kirkwood_formula_intro}) corresponds to (\ref{eq:stokes_fric_intro}) for a viscous incompressible fluid at low Reynolds number, as confirmed by Zwanzig~\cite{Zwanzig} on the basis of fluctuating hydrodynamics~\cite{Landau-LifshitzFluid,Schmitz} and Fax\'en's law~\cite{Kim-Karrila} in the case of the stick boundary condition.
Note that the linear response formula (\ref{eq:kirkwood_formula_intro}) is effective because the Reynolds number is low.

We now consider the derivation of Stokes' law (\ref{eq:stokes_fric_intro}) from Kirkwood's formula (\ref{eq:kirkwood_formula_intro}) without explicitly employing the hydrodynamic equations.
According to the Green--Kubo formula~\cite{Green1954}, the viscosity of the fluid, $\eta$, can be expressed in terms of the stress correlation in the bulk of the fluid as
\begin{equation}
 \eta = \frac{1}{k_{\rm B}T} \int_{0}^{\tau} dt\; \int d^{3} \bm{r} \ave{\sigma^{xy}(\bm{r}_{0},0) \sigma^{xy}(\bm{r},t)}_{\rm eq} ,
  \label{eq:green_kubo_intro}
\end{equation}
where $\sigma^{xy}(\bm{r},t)$ is the $x$-component of the force on the unit area perpendicular to the $y$-axis at position $\bm{r}$ and time $t$ (this is the $xy$ component of the stress tensor $\overleftrightarrow{\bm{\sigma}}$), and $\bm{r}_0$ is an arbitrary position in the bulk of the fluid.
Note that $\tau$ is much larger than the correlation time of $\sigma^{xy}$, and much smaller than the relaxation time of the momentum density field of the fluid.
Precisely speaking, the condition of this $\tau$ is different from that of $\tau$ in Kirkwood's formula, but we assume that there exists some $\tau$ that satisfies both conditions.
Because the right-hand side of (\ref{eq:kirkwood_formula_intro}) can be expressed in terms of the stress correlation on the surface of the sphere, we can obtain Stokes' law (\ref{eq:stokes_fric_intro}) from Kirkwood's formula (\ref{eq:kirkwood_formula_intro}) and the Green--Kubo formula (\ref{eq:green_kubo_intro}) if the surface stress correlation is related to the bulk stress correlation.

The main contribution of this paper is to establish the connection between the bulk and surface stress fluctuations.
The basic concept is simple.
The probability density of surface stress fluctuations is obtained from the probability density of bulk stress fluctuations by integrating out the other degrees of freedom.
This procedure can be conducted in an elegant manner with the aid of large deviation theory.

The remainder of this paper is organized as follows.
In Sect.~\ref{sec:model}, we explain the setup of our model, and we review Kirkwood's formula from the Hamiltonian description of particle systems in Sect.~\ref{sec:kirkwood}.
We derive the probability density of bulk stress fluctuations under two phenomenological assumptions in Sect.~\ref{sec:fluctuation}.
We then express the probability density of surface stress fluctuations by integrating the probability density of bulk stress fluctuations.
We apply a saddle-point method to this expression, and obtain the exact form of the probability density of surface stress fluctuations.
By combining the obtained expression with Kirkwood's formula, we derive Stokes' law.
These highlights are presented in Sect.~\ref{sec:stokes}.
The final section is devoted to a brief summary and some concluding remarks.

Throughout this paper, the superscripts $a,b$ and $\alpha ,\beta$ represent the indices in Cartesian coordinates $(x,y,z)$ and spherical coordinates $(r,\theta,\varphi)$, respectively, where $(x,y,z)=(r\sin\theta\cos\varphi, r\sin\theta\sin\varphi, r\cos\theta)$ with $r\geq 0, 0\leq \theta \leq \pi, 0\leq \varphi \leq 2\pi$.
In addition, we employ Einstein's summation convention for repeated indices appearing in one term.

\section{Model}\label{sec:model}

We provide a three-dimensional mechanical description of our setup.
The system consists of $N$ bath particles of mass $m$ and radius $r_{\rm bp}$, and one macroscopic sphere of mass $M$ and radius $\mathcal{R}$ in a cube of side length $L$.
We assume that $M$ and $\mathcal{R}$ are much larger than $m$ and $r_{\rm bp}$, respectively.
For simplicity, periodic boundary conditions are assumed.
Let $(\bm{r}_{i}, \bm{p}_{i})$ $(1\leq i\leq N)$ be the position and momentum of the $i$th bath particle, and $(\bm{R}, \bm{P})$ be those of the sphere.
A collection of the positions and momenta of all particles is denoted by $\Gamma = (\bm{r}_{1},\bm{p}_{1},\dots ,\bm{r}_{N},\bm{p}_{N},\bm{R},\bm{P})$, which represents the microscopic state of the system.

The Hamiltonian of the system is given by 
\begin{equation}
 H(\Gamma) = \sum_{i=1}^{N} \left[ \frac{\vert \bm{p}_{i}\vert^{2}}{2m} + \sum_{j>i} \Phi_{\rm int}(\vert\bm{r}_{i}-\bm{r}_{j}\vert ) + \Phi_{\rm sp}(\vert \bm{r}_{i}-\bm{R}\vert )\right] + \frac{\vert\bm{P}\vert^2}{2M} ,
  \label{eq:Hamiltonian}
\end{equation}
where $\Phi_{\rm int}$ is a short-range interaction potential between two bath particles, and $\Phi_{\rm sp}$ is that between a bath particle and the sphere.
We assume
\begin{align}
 \Phi_{\rm sp}(\chi ) &\to \infty, \quad \; \text{as } \chi \to \mathcal{R}+r_{\rm bp},
 \label{eq:Phi_sp_1}
 \\
 \Phi_{\rm sp}(\chi ) &=0, \qquad\text{for } \chi \geq \mathcal{R}+r_{\rm bp} + \xi_0,
 \label{eq:Phi_sp_2}
\end{align}
where $\xi_0 \simeq r_{\rm bp}$.
$\Gamma_{t}$ denotes the solution of the Hamiltonian equations at time $t$ for any state $\Gamma$ at $t=0$.
In this setup, the energy is conserved, i.e.,
\begin{equation}
 H(\Gamma_{t})=H(\Gamma),
  \label{eq:energy_conservation}
\end{equation}
and Liouville's theorem
\begin{equation}
 \left\vert \frac{\rd \Gamma_{t}}{\rd \Gamma }\right\vert = 1
  \label{eq:Liouville}
\end{equation}
holds.
The total force acting on the sphere is given by
\begin{align}
 \bm{F}(\Gamma) &= -\frac{\rd H(\Gamma)}{\rd \bm{R}}
 \notag
 \\
 &= - \sum_{i=1}^{N} \frac{\rd \Phi_{\rm sp}(\vert \bm{r}_{i}-\bm{R}\vert )}{\rd \bm{R}}.
 \label{eq:def_force}
\end{align}
For convenience, we abbreviate $\bm{F}(\Gamma_{t})$ as $\bm{F}_{t}$.
The equation of motion for the sphere is written as $\rd_{t} \bm{P}_{t} = \bm{F}_{t}$.

We assume that the system is initially in equilibrium at temperature $T$.
Then, the initial probability density of $\Gamma$ is given by
\begin{equation}
 f_{\rm eq}(\Gamma ) = \exp \left[ - \frac{H(\Gamma) -\Psi_{\rm eq}}{k_{\rm B}T} \right] ,
  \label{eq:dist_eq}
\end{equation}
where $\Psi_{\rm eq}$ is the normalization constant.

\section{Kirkwood's formula}\label{sec:kirkwood}

\subsection{Derivation}

To consider the relaxation of the momentum of the sphere, we apply an impulsive force $M\bm{V}\delta(t)$ to the sphere when the system is in equilibrium at temperature $T$.
Then, the probability density just after $t=0$ is represented by
\begin{equation}
 f_{0}(\Gamma ;\bm{V}) = \exp \left[ - \frac{H(\Gamma) - \bm{P}\cdot \bm{V} -\Psi (\bm{V})}{k_{\rm B}T} \right] ,
  \label{eq:dist_0}
\end{equation}
where $\Psi (\bm{V})$ is the normalization constant and $f_{0}(\Gamma ;\bm{0})=f_{\rm eq}(\Gamma)$.
Using (\ref{eq:energy_conservation}) and (\ref{eq:Liouville}), the probability density at time $t$ is obtained as
\begin{align}
 f_{t}(\Gamma) &= f_{0}(\Gamma_{-t};\bm{V})
 \notag
 \\
 &= \exp \left[ - \frac{H(\Gamma) - \bm{P}_{-t}\cdot \bm{V} -\Psi (\bm{V})}{k_{\rm B}T} \right] .
  \label{eq:dist_t}
\end{align}
In the following, the expectation values with respect to $f_{\rm 0}(\Gamma ;\bm{V})$ and $f_{t}(\Gamma)$ are denoted by $\aves{\; \cdot \; }_{\rm 0}^{\bm{V}}$ and $\aves{\; \cdot \; }_{t}$, respectively.
In particular, $\aves{\; \cdot \;}_{0}^{\bm{0}}$ corresponds to $\aves{\; \cdot \;}_{\rm eq}$.

We derive the exact formula for the friction coefficient for any $\bm{V}$.
Using (\ref{eq:dist_t}) and $\rd_{t}\bm{P}_{-t}=-\bm{F}_{-t}$, we obtain
\begin{align}
 f_{t}(\Gamma) &= f_{0}(\Gamma ;\bm{V}) + \int_{0}^{t} ds\; \rd_s f_{s}(\Gamma)
 \notag
 \\
 &= f_{0}(\Gamma ;\bm{V}) - \int_{0}^{t} ds\; f_{s}(\Gamma) \frac{\bm{F}_{-s}\cdot \bm{V}}{k_{\rm B}T}.
 \label{eq:identity_dist}
\end{align}
In addition, because $\bm{F}$ is independent of the momenta, we obtain
\begin{equation}
 \ave{\bm{F}}_{0}^{\bm{V}} = \ave{\bm{F}}_{\rm eq} = \bm{0}.
  \label{eq:force_eq}
\end{equation}
Then, (\ref{eq:identity_dist}) and (\ref{eq:force_eq}) lead to
\begin{align}
 \ave{F^{a}}_{t} & = -\frac{1}{k_{\rm B}T} \int d\Gamma \int_{0}^{t}ds \; f_{s}(\Gamma ) F^{a}F^{b}_{-s} V^{b}
 \notag
 \\
 &= -\frac{1}{k_{\rm B}T} \int_{0}^{t} ds \int d\Gamma_{-s} \; f_{0}(\Gamma_{-s} ;\bm{V}) F^{a}F^{b}_{-s} V^{b}
 \notag
 \\
 &= - \gamma_{t}^{ab}(\bm{V}) V^{b}
 \label{eq:force_identity}
\end{align}
with
\begin{equation}
 \gamma_{t}^{ab}(\bm{V}) \equiv \frac{1}{k_{\rm B}T} \int_{0}^{t} ds\; \ave{F_s^{a}F^{b}}_{0}^{\bm{V}},
  \label{eq:t_dep_exact_friction}
\end{equation}
which is valid for any $\bm{V}$.
Note that $V^{b}$ is not equal to $P^{b}_{t}/M$ in (\ref{eq:force_identity}), and thus the time dependence of $\aves{F^{a}}_{t}$ is described by that of $\gamma_{t}^{ab}(\bm{V})$.
Furthermore, in general, $\aves{F^{a}}_{t}$ is a nonlinear function of $\bm{V}$.

Next, we focus on the linear response regime.
The dependence of $\gamma_{t}^{ab}$ on $\bm{V}$ is neglected.
Within this regime, we can rewrite (\ref{eq:t_dep_exact_friction}) as
\begin{equation}
 \gamma_{t}^{ab}(\bm{0}) = \gamma_{t} \delta^{ab}
\end{equation}
with
\begin{align}
 \gamma_{t} &= \frac{1}{k_{\rm B}T} \int_{0}^{t} ds\; \ave{F_s^{z}F^{z}}_{\rm eq}
 \notag
 \\
 &= \frac{1}{3k_{\rm B}T}\int_{0}^{t}ds\; \ave{F_s^{a}F^{a}}_{\rm eq},
  \label{eq:kirkwood}
\end{align}
where we have employed the isotropic property of the system in equilibrium.
The expression (\ref{eq:kirkwood}) is the linear response formula for the friction coefficient, first derived by Kirkwood \cite{Kirkwood}.

\subsection{Time dependence of $\gamma_{t}^{ab}(\bm{V})$}
We denote the correlation time of $\bm{F}_{t}$ and the relaxation time of $\bm{P}$ by $\tau_{\rm micro}$ and $\tau_{\rm macro}$, respectively.
We assume the separation of time scales represented by $\tau_{\rm micro} \ll \tau_{\rm macro}$.
Here, (\ref{eq:force_eq}) and $\aves{\bm{P}}^{\bm{V}}_{0}=M\bm{V}$ lead to
\begin{align}
 \ave{P^{a}F^{b}}^{\bm{V}}_{0} &= \int d\Gamma \; f_{0}(\Gamma ; \bm{V}) \left( P^{a}-MV^{a}\right) F^{b}
 \notag
 \\
 &= 0.
  \label{eq:cor_PF_ini}
\end{align}
Using $\rd_{t} \bm{P}_{t}=\bm{F}_{t}$ and (\ref{eq:cor_PF_ini}), we can rewrite (\ref{eq:t_dep_exact_friction}) as
\begin{align}
 \gamma_{t}^{ab}(\bm{V}) &= \frac{1}{k_{\rm B}T} \left[ \ave{P_{t}^{a}F^{b}}^{\bm{V}}_{0} - \ave{P^{a}F^{b}}^{\bm{V}}_{0} \right]
  \notag
  \\
  &= \frac{1}{k_{\rm B}T} \ave{P_{t}^{a}F^{b}}^{\bm{V}}_{0}.
  \label{eq:t_dep_exact_friction_2}
\end{align}
Thus, we find $\gamma_{t}^{ab}(\bm{V})=0$ for any $\bm{V}$ when $t \gg \tau_{\rm macro}$, because the correlation between $P^{a}$ and $F^{b}$ is considered to take about $\tau_{\rm macro}$.
Furthermore, by considering (\ref{eq:t_dep_exact_friction}), we obtain $\gamma_{0}^{ab}(\bm{V})= 0$ for any $\bm{V}$.
Keeping these two limiting cases in mind, we conjecture the following time dependence of $\gamma_{t}^{ab}(\bm{V})$.
$\gamma_{t}^{ab}(\bm{V})$ increases when $0<t\simeq \tau_{\rm micro}$.
After that, $\gamma_{t}^{ab}(\bm{V})$ remains constant when $\tau_{\rm micro}\ll t\ll \tau_{\rm macro}$.
Eventually, $\gamma_{t}^{ab}(\bm{V})$ gradually tends to zero when $t\simeq \tau_{\rm macro}$.
By recalling (\ref{eq:t_dep_exact_friction}), we can express this behavior as
\begin{equation}
 \ave{F_t^{a}F^{b}}_{0}^{\bm{V}} 
 \begin{cases}
  > 0, & \text{for } 0< t \simeq \tau_{\rm micro},
  \\
  = 0, & \text{for } \tau_{\rm micro}\ll t \ll \tau_{\rm macro},
  \\
  < 0, & \text{for } \tau\simeq \tau_{\rm macro}.
 \end{cases}
\end{equation}
In the linear response regime, we denote the constant value of $\gamma_{t}$ during the time interval $\tau_{\rm micro}\ll t\ll \tau_{\rm macro}$ as $\gamma$.
Using this particular value of $\gamma$, $\tau_{\rm macro}$ can be expressed as $M/\gamma$.
This $\gamma$ is the linear friction coefficient.

\subsection{Other expressions}

We hereafter focus on the equilibrium case in which $\Gamma$ is chosen according to the canonical ensemble $f_{\rm eq}(\Gamma)$, and consider a finite time interval $[0, \tau]$ that satisfies $\tau_{\rm micro}\ll \tau\ll \tau_{\rm macro}$. 
Because the motion of the sphere can be ignored up to $\tau$, we assume that the center of the sphere is fixed at the origin.
Throughout this paper, for any physical quantity $A(\Gamma)$, we define the time-averaged quantity by
\begin{equation}
\bar{A}(\Gamma) \equiv \frac{1}{\tau}\int_{0}^{\tau} dt\; A(\Gamma_t).
\end{equation}
Then, we can rewrite (\ref{eq:kirkwood}) as
\begin{equation}
\gamma = \frac{\tau}{2k_{\rm B}T} \ave{\left( \bar{F}^{z} \right)^2}_{\rm eq}.
\label{eq:kirkwood_square}
\end{equation}
In this derivation, it should be noted that $\aves{F^{z}_t F^{z}_s}_{\rm eq}$ is a function of $\vert t-s\vert$, and that $\aves{F^{z}_t F^{z}}_{\rm eq} = 0$ when $\tau_{\rm micro} \ll t \ll \tau_{\rm macro}$.

We can also express the force $F^{a}_{t}$ by the surface integration of a stress $\sigma^{ab}_{\rm sp}$:
\begin{equation}
 F^{a}(\Gamma) = \mathcal{R}^2 \int d\Omega \; n^b \sigma^{ab}_{\rm sp}(\mathcal{R},\Omega ;\Gamma),
  \label{eq:F_and_sigma}
\end{equation}
where $\Omega$ is a solid angle and $\bm{n}\equiv ( \sin \theta \cos \varphi , \sin \theta \sin \varphi , \cos \theta)$.
By direct calculation, we find 
\begin{align}
{\sigma}^{ab}_{\rm sp}(\bm{r};\Gamma) &= -\sum_{i=1}^{N} \left[ -\frac{\rd \Phi_{\rm sp}(\vert \bm{r}_{i}-\bm{R}\vert )}{\rd R^{a}}\right] \left( R^{b}-r_{i}^{b}\right) D(\bm{r};\bm{r}_{i},\bm{R})
\label{eq:micro_stress:sp}
\end{align}
with
\begin{equation}
 D(\bm{r};\bm{r}_{i},\bm{R}) = \int_{0}^{1} d\xi\; \delta \left( \bm{r}-\bm{r}_{i}-(\bm{R}-\bm{r}_{i})\xi\right).
\end{equation}
Indeed, by substituting (\ref{eq:micro_stress:sp}) into (\ref{eq:F_and_sigma}), we obtain (\ref{eq:def_force}), where we have used the divergence theorem and
\begin{equation}
 (R^{b}-r^{b}_{i}) \rd^{b} D(\bm{r};\bm{r}_{i},\bm{R}) = \delta (\bm{r}-\bm{r}_{i}) - \delta (\bm{r}-\bm{R}).
\end{equation}
Furthermore, the stress tensor can be defined in spherical coordinates by
\begin{equation}
 \begin{bmatrix}
  \sigma_{\rm sp}^{xx} & \sigma_{\rm sp}^{xy} & \sigma_{\rm sp}^{xz} \\[3pt]
  \sigma_{\rm sp}^{yx} & \sigma_{\rm sp}^{yy} & \sigma_{\rm sp}^{yz} \\[3pt]
  \sigma_{\rm sp}^{zx} & \sigma_{\rm sp}^{zy} & \sigma_{\rm sp}^{zz}
 \end{bmatrix}
 = Q
 \begin{bmatrix}
  \sigma_{\rm sp}^{rr} & \sigma_{\rm sp}^{r\theta} & \sigma_{\rm sp}^{r\varphi} \\[2pt]
  \sigma_{\rm sp}^{\theta r} & \sigma_{\rm sp}^{\theta\theta} & \sigma_{\rm sp}^{\theta\varphi} \\[2pt]
  \sigma_{\rm sp}^{\varphi r} & \sigma_{\rm sp}^{\varphi\theta} & \sigma_{\rm sp}^{\varphi\varphi}
 \end{bmatrix}
 Q^{\rm T}
\end{equation}
with
\begin{equation}
 Q=
 \begin{bmatrix}
  \; \sin\theta \cos \varphi\; & \; \cos\theta \cos\varphi\; & \; -\sin\varphi\; \\
  \sin\theta \sin \varphi & \cos\theta \sin \varphi & \cos \varphi \\
  \cos\theta & -\sin\theta & 0
 \end{bmatrix},
\end{equation}
where $Q^{\rm T}$ denotes the transpose of $Q$.
We then obtain 
\begin{equation}
 F^{z} (\Gamma) = \mathcal{R}^2 \int d\Omega \; \left[ \cos \theta\; \sigma^{rr}_{\rm sp}(\mathcal{R},\Omega;\Gamma) -\sin \theta\; \sigma^{\theta r}_{\rm sp}(\mathcal{R},\Omega;\Gamma) \right] .
  \label{eq:force_stress_sp}
\end{equation}

Here, it is convenient to define the stress $\sigma_{*}(\Gamma)$ that represents the $z$-component of the force per unit area on the surface of the sphere.
From (\ref{eq:force_stress_sp}) and $F^{z}(\Gamma)=4\pi \mathcal{R}^2 \sigma_{*}(\Gamma)$, we have
\begin{equation}
 \sigma_{*}(\Gamma) = \frac{1}{4\pi} \int d\Omega \; \left[ \cos \theta\; \sigma^{rr}_{\rm sp}(\mathcal{R},\Omega; \Gamma) - \sin \theta\; \sigma^{\theta r}_{\rm sp}(\mathcal{R},\Omega; \Gamma)\right] .
\label{eq:stress_fix}
\end{equation}
The formula (\ref{eq:kirkwood_square}) is rewritten as
\begin{equation}
\gamma = \frac{(4\pi \mathcal{R}^2)^2 \tau}{2k_{\rm B}T} \ave{( \bar{\sigma}_{*} )^2 }_{\rm eq}.
\label{eq:kirkwood_stress}
\end{equation}
The linear friction coefficient is expressed in terms of the correlation of the time-averaged stress at the surface, $\bar{\sigma}_{*}$.

\section{Stress fluctuations in equilibrium viscous fluids}
\label{sec:fluctuation}

Our goal is to derive Stokes' law (\ref{eq:stokes_fric_intro}) from Kirkwood's formula (\ref{eq:kirkwood_stress}) without explicitly employing the Stokes equations that are a linearized form of the Navier--Stokes equations in the low Reynolds number limit.
Stokes' law (\ref{eq:stokes_fric_intro}) contains the viscosity of the fluid $\eta$.
Instead of introducing $\eta$ as a parameter of the Stokes equations, we define $\eta$ from the stress fluctuations in the bulk of the fluid by the Green--Kubo formula (\ref{eq:green_kubo_intro}).
Thus, to derive Stokes' law, we must relate the bulk stress correlation to the surface stress correlation.
In this section, we derive the probability density of coarse-grained time-averaged stress fluctuations in equilibrium viscous fluids. 

\subsection{The Green--Kubo formula}
\label{sec:Green-Kubo}

We first consider the microscopic expression of the stress $\sigma^{ab}(\bm{r};\Gamma)$ in the bulk of the equilibrium fluid.
For the momentum density of the bath particles 
\begin{equation}
{\Pi}^{a} (\bm{r};\Gamma) \equiv \sum_{i=1}^{N} p_{i}^{a}\delta (\bm{r}-\bm{r}_{i}),
\end{equation}
${\sigma}^{ab}(\bm{r};\Gamma)$ is defined to satisfy the following continuity equation:
\begin{equation}
\rd_{t} {\Pi}^{a} (\bm{r};\Gamma_{t}) = \rd^{b} \left[ {\sigma}^{ab}(\bm{r};\Gamma_{t}) + {\sigma}^{ab}_{\rm sp}(\bm{r};\Gamma_{t})\right] .
\end{equation}
Note that the stress ${\sigma}^{ab}(\bm{r};\Gamma)$ corresponds to the force per unit area ${\sigma}^{ab}(\bm{r};\Gamma) n^b$ exerted on a virtual surface described by the unit vector $n^b$ from the outer side, where the unit vector is perpendicular to the surface and directed toward the outer region.
${\sigma}^{ab}_{\rm sp}$ was given in (\ref{eq:micro_stress:sp}).
Here, ${\sigma}^{ab}(\bm{r};\Gamma)$ is calculated as 
\begin{align}
{\sigma}^{ab}(\bm{r};\Gamma) &= -\sum_{i=1}^{N} \frac{ p_{i}^{a} p_{i}^{b}}{m} \delta \left( \bm{r}-\bm{r}_{i}\right)  - \sum_{i<j} \left[ -\frac{\rd \Phi_{\rm int}(\vert\bm{r_{i}-\bm{r}_{j}} \vert )}{\rd \vert\bm{r_{i}-\bm{r}_{j}} \vert} \right] \frac{\left( r_{i}^{a} - r_{j}^{a}\right)\left( r_{i}^{b} - r_{j}^{b}\right) }{\vert\bm{r_{i}-\bm{r}_{j}} \vert} D(\bm{r};\bm{r}_{i},\bm{r}_{j}).
\label{eq:micro_stress}
\end{align}
It can be directly confirmed that 
\begin{equation}
{\sigma}^{ab}(\bm{r};\Gamma) ={\sigma}^{ba}(\bm{r};\Gamma) .
\label{symmetric-tensor}
\end{equation}
Note that, when there are interactions between bath particles across the periodic boundary, the right-hand side of (\ref{eq:micro_stress}) must be modified to take a non-zero value on the shortest line connecting two interacting bath particles.

By employing the microscopic expression of the stress, the Green--Kubo formula (\ref{eq:green_kubo_intro}) is precisely expressed as
\begin{equation}
\eta = \frac{1}{k_{\rm B}T} \int_{0}^{\tau} dt\; \int d^{3} \bm{r} 
\ave{\sigma^{xy}(\bm{r}_{0};\Gamma) \sigma^{xy}(\bm{r};\Gamma_t)}_{\rm eq},
\label{eq:green_kubo_intro:precise}
\end{equation}
where $\tau$ is chosen such that $\tau_{\rm micro}^{\rm fluid} \ll \tau \ll \tau_{\rm macro}^{\rm fluid}$.
$\tau_{\rm micro}^{\rm fluid}$ is the correlation time of $\sigma^{xy}$, and $\tau_{\rm macro}^{\rm fluid}$ is the relaxation time of $\bm{\Pi}$.
In general, this $\tau$ is different from that in Kirkwood's formula (\ref{eq:kirkwood_stress}).
Here, we assume that the same value of $\tau$ can be chosen in the two formulas, and use the same notation.
One may notice the long-time tail of the stress correlation, which apparently breaks the separation of time scales.
We assume that this long-time tail only appears in the regime $t\gtrsim \tau$, because it originates from long wavelength fluctuations of locally conserved quantities.
Therefore, we continue the argument without considering the long-time tail problem.

\subsection{Macroscopic fluctuation theory}

Let $\xi_{\rm micro}$ be the largest length scale appearing in the molecular description, and $\xi_{\rm macro}$ be the minimum length characterizing macroscopic behavior.
In the case of viscous liquids, $\xi_{\rm micro}$ may be estimated as $r_{\rm bp}$, and $\xi_{\rm macro}$ is given by $\mathcal{R}$.
Because $\xi_{\rm micro} \ll \xi_{\rm macro}$, we can choose $\Lambda$ to satisfy $\xi_{\rm micro} \ll \Lambda \ll \xi_{\rm macro}$. 
We then define the coarse-grained time-averaged stresses by
\begin{equation}
\hat \sigma^{ab}(\bm{r}) = \frac{1}{\Lambda^3}\int_{\mathcal{V}({\bm{r}})} 
d^3 \bm{r}'\; \bar{\sigma}^{ab}(\bm{r}')
\end{equation}
with $\mathcal{V}(\bm{r}) \equiv \prod_{a}\left[ r^{a}-\Lambda /2, r^{a}+\Lambda /2\right]$.
In terms of $\hat \sigma^{ab}(\bm{r})$, the Green--Kubo formula (\ref{eq:green_kubo_intro:precise}) is expressed as
\begin{equation}
\eta = \frac{\tau \Lambda^3 }{2k_{\rm B}T} \ave{ (\hat \sigma^{xy}(\bm{r}))^2  }_{\rm eq}.
\label{eq:green_kubo_intro:c-g}
\end{equation}
We also note that 
\begin{equation}
\frac{\tau \Lambda^3 }{2k_{\rm B}T} \ave{ \hat{\sigma}^{xy}( \bm{r})\hat{\sigma}^{xy}(\bm{r}')}_{\rm eq} \simeq 0
\label{eq:green_kubo_intro:c-g-off}
\end{equation}
for $\vert\bm{r}-\bm{r}'\vert \ge \Lambda$, because the correlation length of the stress field is on the order of $\xi_{\rm micro}$.
The relations (\ref{eq:green_kubo_intro:c-g}) and (\ref{eq:green_kubo_intro:c-g-off}) can be summarized as
\begin{equation}
\ave{ \hat{\sigma}^{xy}({\bm{r}}) \hat{\sigma}^{xy}({\bm{r}'})}_{\rm eq} =\frac{2k_{\rm B}T \eta}{\tau \Lambda^3} \mathcal{S}(\vert {\bm{r}}-{\bm{r}'}\vert ),
\label{eq:green_kubo_intro:c-g-sum}
\end{equation}
where $\mathcal{S}(d)=1$ for  $d \ll \Lambda$ and $\mathcal{S}(d)\simeq 0$ for $d\ge \Lambda$.

We now formulate a macroscopic fluctuation theory for the coarse-grained time-averaged stresses $\hat{\sigma}^{ab}$.
Because $\hat \sigma^{ab}({\bm{r}})$ are obtained by integrating the microscopic stress over the region $\mathcal{V}(\bm{r})$, it is reasonable, by considering the central limit theorem, to expect these $\hat \sigma^{ab}({\bm{r}})$ to obey a Gaussian distribution.
Here, we adapt the continuum description with the space mesh (ultraviolet cutoff) $\Lambda$.
In this description, $\mathcal{S}(\vert \bm{r}-\bm{r}' \vert) /\Lambda^3 \to \delta (\bm{r}-\bm{r}')$, and thus (\ref{eq:green_kubo_intro:c-g-sum}) is rewritten as
\begin{equation}
\ave{ \hat{\sigma}^{xy}({\bm{r}}) \hat{\sigma}^{xy}({\bm{r}'})}_{\rm eq}
= \frac{2k_{\rm B}T \eta}{\tau} \delta({\bm{r}}-{\bm{r}'}).
\label{eq:green_kubo_intro:c-g-2}
\end{equation}

Next, we consider the statistical properties of the stresses.
We start with the following decomposition of $\hat{\sigma}^{ab}$ into traceless and trace parts:
\begin{equation}
\hat{\sigma}^{ab}(\bm{r}) = - \hat{p}(\bm{r}) \delta^{ab} + \hat{s}^{ab}(\bm{r})
\label{eq:decomp_stress}
\end{equation}
with
\begin{equation}
\hat{p}(\bm{r}) \equiv  -\frac{\hat{\sigma}^{xx}(\bm{r})
+\hat{\sigma}^{yy}(\bm{r})+\hat{\sigma}^{zz}(\bm{r})}{3},
\label{eq:pres_stress}
\end{equation}
where $\hat{p}$ and $\hat{s}^{ab}$ are called the mean pressure and the stress deviator tensor, respectively.
Because $\hat{s}^{ab}$ is traceless, we have
\begin{equation}
\hat{s}^{xx}(\bm{r})+\hat{s}^{yy}(\bm{r})+\hat{s}^{zz}(\bm{r})=0.
\label{eq:noise_traceless}
\end{equation}
Using the isotropic property and the assumption that the correlation length of the stress fluctuations is much less than $\Lambda$, we can express
\begin{equation}
 \ave{\hat{s}^{ab}(\bm{r})\hat{s}^{a'b'}(\bm{r}')}_{\rm eq} = \left[ B_1 \delta^{aa'}\delta^{bb'} + B_2 \delta^{ab'}\delta^{ba'} + B_3 \delta^{ab}\delta^{a'b'}\right] \delta (\bm{r}-\bm{r}'),
\end{equation}
where the constants $B_1$, $B_2$, and $B_3$ are determined below.
By recalling (\ref{symmetric-tensor}), we obtain $\hat{\sigma}^{ab}(\bm{r})=\hat{\sigma}^{ba}(\bm{r})$, and then we find that
\begin{equation}
 \ave{\hat{s}^{xy}(\bm{r})\hat{s}^{xy}(\bm{r}')}_{\rm eq} = \ave{\hat{s}^{yx}(\bm{r})\hat{s}^{xy}(\bm{r}')}_{\rm eq},
\end{equation}
which leads to $B_1=B_2$.
Using (\ref{eq:noise_traceless}), we also have
\begin{equation}
 \ave{\left( \hat{s}^{xx}(\bm{r}) + \hat{s}^{yy}(\bm{r}) + \hat{s}^{zz}(\bm{r})\right) \left( \hat{s}^{xx}(\bm{r}') + \hat{s}^{yy}(\bm{r}') + \hat{s}^{zz}(\bm{r}')\right)}_{\rm eq} = 0,
\end{equation}
which gives $B_3=-2B_{1}/3$.
Finally, the Green--Kubo formula (\ref{eq:green_kubo_intro:c-g-2}) leads to
\begin{equation}
 B_{1} = \frac{2k_{\rm B}T\eta}{\tau}.
\end{equation}
We summarize these relations as
\begin{align}
\ave{\hat{s}^{ab}(\bm{r})\hat{s}^{a'b'}(\bm{r}')}_{\rm eq} = \frac{2k_{\rm B}T\eta}{\tau} \Delta^{aba'b'}\delta (\bm{r}-\bm{r}')
\label{eq:stress_noise_Car}
\end{align}
with
\begin{equation}
\Delta^{aba'b'} = \delta^{aa'}\delta^{bb'} + \delta^{ab'}\delta^{ba'} - \frac{2}{3}\delta^{ab}\delta^{a'b'}.
\label{eq:iso_tensor_4}
\end{equation}

Condition (\ref{eq:stress_noise_Car}) is necessary, but the statistical distribution of $\hat{\sigma}^{ab}$ is not completely determined.
The problem is to derive the statistical property of $\hat p$.
Because we are studying incompressible fluids, the pressure fluctuation arising from the density fluctuations need not be taken into account.
Rather, we assume that $\hat p$ is determined from the balance of time-averaged forces in each region, which is expressed as $\rd^{b} \hat{\sigma}^{ab}(\bm{r})=0$.
We express this in coordinate-free form as
\begin{equation}
\nabla \cdot \overleftrightarrow{\hat{\bm{\sigma}}}(\bm{r}) = \bm{0}.
\label{eq:stress_balance}
\end{equation}
It is obvious that the expectation values of the stresses satisfy the balance condition in the equilibrium cases. However, assumption (\ref{eq:stress_balance}) means that fluctuating stresses are already balanced in the macroscopic fluctuation theory.

\subsection{Probability density of stresses}

For later convenience, we express  $\Delta^{aba'b'} =\Delta^{ij}$ with $g(ab)=i$ and $g(a'b')=j$, where $g(xx)=1$, $g(yy)=2$, $g(zz)=3$, $g(xy)=g(yx)=4$, $g(yz)=g(zy)=5$, and $g(zx)=g(xz)=6$. 
That is, $\Delta$ is interpreted as a $6 \times 6$ matrix. 
(We use the same symbol $\Delta$ without confusion.)
Explicitly, the matrix $\Delta$ is expressed as
\begin{equation}
\Delta = 
 \begin{bmatrix}
  4/3 & -2/3 & -2/3 & \; 0\; & \; 0\; & \; 0\; \\
  -2/3 & 4/3 & -2/3 & 0 & 0 & 0 \\
  -2/3 & -2/3 & 4/3 & 0 & 0 & 0 \\
  0 & 0 & 0 & 1 & 0 & 0 \\
  0 & 0 & 0 & 0 & 1 & 0 \\
  0 & 0 & 0 & 0 & 0 & 1
 \end{bmatrix}.
\end{equation}
As the probability density of $\hat{s}^{ab}$ is expressed using the inverse of $\Delta$, we calculate the $k$th eigenvalue $\epsilon_k$ and corresponding eigenvector  $\phi_k^j$ defined by
\begin{equation}
\sum_{j=1}^{6} \Delta^{ij}\phi_k^j=\epsilon_k \phi_k^i.
\end{equation}
We find that
\begin{equation}
 \epsilon_{k} =
  \begin{cases}
   0 & \text{for } k=1 \\
   2 & \text{for } k=2,3 \\
   1 & \text{for } k=4,5,6
  \end{cases}.
\end{equation}
The zero-eigenvector is calculated as $\phi_1^j=1/\sqrt{3}$ for $j=1,2,3$ and $\phi_1^j=0$  for $j=4,5,6$.
Although the matrix $\Delta$ is singular, we can define the pseudo-inverse of $\Delta$ as
\begin{equation}
 \sum_{j=1}^{6} ( \Delta^{-1})^{ij}\phi_1^j= 0,
\end{equation}
and 
\begin{equation}
\sum_{j=1}^{6} (\Delta^{-1})^{ij}\phi_k^j=\epsilon_k^{-1} \phi_k^i
\end{equation}
for $k\ge 2$.
A straightforward calculation yields
\begin{equation}
\Delta^{-1}= 
 \begin{bmatrix}
 1/3 & -1/6 & -1/6 & \; 0\; & \; 0\; & \; 0\; \\
-1/6 & 1/3 & -1/6 & 0 & 0 & 0 \\
-1/6 & -1/6 & 1/3 & 0 & 0 & 0 \\
  0 & 0 & 0 & 1 & 0 & 0 \\
  0 & 0 & 0 & 0 & 1 & 0 \\
  0 & 0 & 0 & 0 & 0 & 1
 \end{bmatrix}.
\end{equation}
We identify $(\Delta^{-1})^{aba'b'}$ with $(\Delta^{-1})^{ij}$ by $i=g(ab)$ and $j=g(a'b')$.

Now, we write the probability density of $\{ \hat{p}(\bm{r}), \hat{s}^{ab}(\bm{r})\}$ such that the statistical properties of the macroscopic stress fields $\{ \hat{p}(\bm{r}), \hat{s}^{ab}(\bm{r})\}$ are reproduced.
Because the macroscopic stress fields obey the Gaussian distribution on the restricted configuration space given by (\ref{eq:noise_traceless}) and (\ref{eq:stress_balance}), the expression is 
\begin{align}
\mathcal{P}\left( \{ \hat{p}(\bm{r}), \hat{s}^{ab}(\bm{r})\} \right) 
&= C \exp \left[ - \frac{\tau}{4k_{\rm B}T\eta } 
\int d^3\bm{r}\; 
\hat{s}^{{a}{b}}(\bm{r})
(\Delta^{-1})^{{a}{b}{a}'{b}'}\hat{s}^{{a}'{b}'}(\bm{r})\right]
\notag
\\
& \qquad \times \prod_{\bm{r}} \delta \Big( \hat{s}^{xx}(\bm{r}) + \hat{s}^{yy}(\bm{r}) + \hat{s}^{zz}(\bm{r}) \Big) \times \prod_{\bm{r}} \prod_{a} \delta \Big( \rd^b \hat{s}^{ab}(\bm{r}) - \rd^{a} \hat{p}(\bm{r}) \Big) .
 \label{eq:bulk_pro_p_s}
\end{align}
where $C$ is the normalization constant. 

Because the traceless condition (\ref{eq:noise_traceless}) can be written as 
\begin{equation}
\phi_1^{g(ab)} \hat{s}^{ab}(\bm{r}) =0,
\label{eq:noise_traceless-vec}
\end{equation}
we find that $ \hat{s}^{{a}{b}}(\Delta^{-1})^{{a}{b}{a}'{b}'}\hat{s}^{{a}'{b}'}$ in (\ref{eq:bulk_pro_p_s}) can be replaced by $\hat{s}^{{a}{b}}(\tilde \Delta^{-1})^{{a}{b}{a}'{b}'}\hat{s}^{{a}'{b}'}$ with 
\begin{equation}
(\tilde \Delta^{-1})^{ij}= (\Delta^{-1})^{ij}+\psi^i \phi_1^j+\psi^j \phi_1^i,
\end{equation}
where $\psi^i$ is an arbitrary vector.
Explicitly, this is written as
\begin{equation}
\tilde \Delta^{-1}=  \Delta^{-1}+
 \begin{bmatrix}
  \; 2q_1\; & \; q_1+q_2\; & \; q_1+q_3\; & \; q_4\; & \; q_5\; & \; q_6\; \\
  q_1+q_2 & 2q_2 & q_2+q_3 & q_4 & q_5 & q_6 \\
  q_1+q_3 & q_2+q_3 & 2q_3 & q_4 & q_5 & q_6 \\
  q_4 & q_4 & q_4 & 0 & 0 & 0 \\
  q_5 & q_5 & q_5 & 0 & 0 & 0 \\
  q_6 & q_6 & q_6 & 0 & 0 & 0
 \end{bmatrix}.
\end{equation}
With the particular choices $q_1=q_2=1/3$, $q_3=-1/6$, and $q_4=q_5=q_6=0$, $\tilde \Delta^{-1}$ takes the simpler form:
\begin{equation}
\tilde  \Delta^{-1} =
 \begin{bmatrix}
  1 & 1/2 & \; 0\; & \; 0\; & \; 0\; & \; 0\; \\
  1/2 & 1 &0 & 0 & 0 & 0 \\
  0 & 0 & 0 & 0 & 0 & 0 \\
  0 & 0 & 0 & 1 & 0 & 0 \\
  0 & 0 & 0 & 0 & 1 & 0 \\
  0 & 0 & 0 & 0 & 0 & 1
 \end{bmatrix}.
 \label{eq:Delta_inv}
\end{equation}
Note that $\tilde  \Delta^{-1} $ is the pseudo-inverse of $\tilde \Delta$ given by 
\begin{equation}
\tilde  \Delta = 
 \begin{bmatrix}
 4/3 & -2/3 & \; 0\; & \; 0\; & \; 0\; & \; 0\; \\
-2/3 & 4/3 &0 & 0 & 0 & 0 \\
  0 & 0 & 0 & 0 & 0 & 0 \\
  0 & 0 & 0 & 1 & 0 & 0 \\
  0 & 0 & 0 & 0 & 1 & 0 \\
  0 & 0 & 0 & 0 & 0 & 1
 \end{bmatrix},
\end{equation}
where $\tilde  \Delta $ is obtained by setting the $(zz)$ column
and the $(zz)$ row to zero in the matrix $\Delta$ . 

The transformation of variables from $\hat{s}^{ab}$ to $\hat{\sigma}^{ab}$ yields
\begin{align}
 \mathcal{P}\left( \{ \hat{p}(\bm{r}), \hat{\sigma}^{ab}(\bm{r})\} \right) &= C' \exp \left[ - \tau \mathcal{I} \left( \{ \hat{p}(\bm{r}), \hat{\sigma}^{ab}(\bm{r})\} \right) \right]
 \notag
 \\
 & \qquad \times \prod_{\bm{r}} \delta \Big( \hat{\sigma}^{xx}(\bm{r}) + \hat{\sigma}^{yy}(\bm{r}) + \hat{\sigma}^{zz}(\bm{r}) + 3\hat{p}(\bm{r}) \Big) 
\delta \Big( \nabla \cdot \overleftrightarrow{\hat{\bm{\sigma}}}(\bm{r})\Big) 
\label{eq:pro_all_stress_eu}
\end{align}
with
\begin{align}
 \mathcal{I}\left( \{ \hat{p}(\bm{r}), \hat{\sigma}^{ab}(\bm{r})\} \right)  =  \frac{1}{4k_{\rm B}T\eta} \int d^3\bm{r}\; \left( \hat{\sigma}^{ab}(\bm{r}) +\hat{p}(\bm{r})\delta^{ab}\right) (\tilde \Delta^{-1})^{aba'b'} \left( \hat{\sigma}^{a'b'}(\bm{r})+\hat{p}(\bm{r}) \delta^{a'b'} \right) ,
\end{align}
where $C'$ is the normalization constant.
Because $\tau \gg \tau_{\rm micro}$, $\mathcal{I}( \{ \hat{p}(\bm{r}),\hat{\sigma}^{ab}(\bm{r})\} )$ corresponds to a large deviation function in probability theory~\cite{Dembo-Zeitouni,Touchette}.

Finally, we consider the stress tensor in spherical coordinates. This is denoted by $\hat \sigma^{\alpha \beta}$, where $\alpha$ and $\beta$ represent $r$, $\theta$, or $\varphi$. 
With a similar method, we obtain 
\begin{align}
 \mathcal{P}\left( \{ \hat{p}(\bm{r}), \hat{\sigma}^{\alpha \beta}(\bm{r})\} \right) &= C'' \exp \left[ - \tau \mathcal{I} \left( \{ \hat{p}(\bm{r}), \hat{\sigma}^{\alpha \beta}(\bm{r})\} \right) \right]
 \notag
 \\
 & \qquad \times \prod_{\bm{r}}\delta \Big( \hat{\sigma}^{rr}(\bm{r}) + \hat{\sigma}^{\theta\theta}(\bm{r}) + \hat{\sigma}^{\varphi\varphi}(\bm{r}) + 3\hat{p}(\bm{r}) \Big) \delta \Big( \nabla \cdot \overleftrightarrow{\hat{\bm{\sigma}}}(\bm{r})\Big) 
  \label{eq:pro_all_stress}
\end{align}
with
\begin{align}
 \mathcal{I}\left( \{ \hat{p}(\bm{r}), \hat{\sigma}^{\alpha \beta}(\bm{r})\right) &= \frac{1}{4k_{\rm B}T\eta } \int_{\mathcal{R}}^{\infty} dr \int_{0}^{\pi} d\theta \int_{0}^{2\pi} d\varphi\; r^2 \sin \theta
 \notag
 \\
 & \qquad \times \left( \hat{\sigma}^{{\alpha}{\beta}}(\bm{r})+\hat{p}(\bm{r})\delta^{{\alpha}{\beta}}\right) (\tilde \Delta^{-1})^{{\alpha}{\beta}{\alpha}'{\beta}'} \left( \hat{\sigma}^{{\alpha}'{\beta}'}(\bm{r}) +\hat{p}(\bm{r})\delta^{{\alpha}'{\beta}'} \right) ,
\label{eq:large_deviation_p_sigma}
\end{align}
where $C''$ is the normalization constant and we have taken the limit $L \to \infty$.
$\tilde \Delta$ is obtained by setting the $(\varphi\varphi)$ column and the $(\varphi\varphi)$ row to zero in $\Delta$ .
Note that the fourth-order isotropic tensor $\Delta^{aba'b'}$ takes the same form as in Cartesian coordinates. 

Because $\mathcal{R}\gg r_{\rm bp}$ and the interaction potential between the sphere and a bath particle depends only on the distance between them, the momentum of the tangential direction at the surface of the sphere is conserved.
This yields the boundary condition 
\begin{equation}
\hat \sigma^{\theta r}(\mathcal{R},\Omega)
= \hat \sigma^{\varphi r}(\mathcal{R},\Omega) = 0.
\label{eq:slip_cond}
\end{equation}

\section{Stress fluctuation at surface}\label{sec:stokes}

Let us return to Kirkwood's formula (\ref{eq:kirkwood_stress}).
We want to calculate the friction constant $\gamma$ from the fluctuation intensity of $\bar{\sigma}_{*}$.
We first connect $\bar{\sigma}_{*}$ with $\hat{\sigma}^{\alpha \beta}$.
By recalling that $\mathcal{R}\gg r_{\rm bp}$ and using (\ref{eq:Phi_sp_1}) and (\ref{eq:micro_stress}), we obtain
\begin{equation}
 \sigma^{ab}(\mathcal{R},\Omega ;\Gamma )=0
\end{equation}
at the surface of the sphere.
From (\ref{eq:Phi_sp_2}) and (\ref{eq:micro_stress:sp}), we obtain
\begin{equation}
 \sigma^{ab}_{\rm sp}(\mathcal{R}+r_{\rm bp}+\xi_0,\Omega;\Gamma ) = 0.
\end{equation}
The continuity of the total coarse-grained time-averaged stresses leads to 
\begin{equation}
 \hat{\sigma}^{ab}(\mathcal{R},\Omega) = \frac{1}{\Lambda^2}\int_{S(\mathcal{R},\Omega)} d\Omega' \; \bar{\sigma}^{ab}_{\rm sp}(\mathcal{R},\Omega'),
  \label{eq:stress_continuity}
\end{equation}
where $S(\mathcal{R},\Omega)$ is the region on the surface of the sphere represented by
\begin{equation}
  \left[ \theta - \frac{\Lambda}{2\mathcal{R}}, \theta + \frac{\Lambda}{2\mathcal{R}}\right] \times \left[ \varphi - \frac{\Lambda}{2\mathcal{R}\sin\theta}, \varphi + \frac{\Lambda}{2\mathcal{R}\sin\theta}\right] .
\end{equation}
Therefore, from (\ref{eq:stress_fix}) and (\ref{eq:stress_continuity}), we find that $\bar{\sigma}_*$ in (\ref{eq:kirkwood_stress}) is connected to $\hat{\sigma}^{\alpha\beta}$ by the relation
\begin{align}
 \bar{\sigma}_{*} &= \frac{1}{4\pi} \int d\Omega \; \left[ \cos \theta\; \hat \sigma^{rr}(\mathcal{R},\Omega) - \sin \theta\; \hat \sigma^{r\theta}(\mathcal{R},\Omega)\right].
\label{eq:stress_fix-con}
\end{align}
The probability density of $\bar{\sigma}_{*}$ is determined from the statistical properties of $\hat{\sigma}^{\alpha\beta}$ in the bulk under conditions (\ref{eq:slip_cond}) and  (\ref{eq:stress_fix-con}).
This relation is formally written as
\begin{equation}
 P(\bar{\sigma}_{*}) = \int_{\bar{\sigma}_{*}\text{:fix}} \mathcal{D}\hat{p} \mathcal{D} \hat{\sigma}^{\alpha \beta} \; \mathcal{P} \left( \{ \hat{p}(\bm{r}), \hat{\sigma}^{\alpha \beta}(\bm{r})\} \right),
\label{eq:pro_sigma_*}
\end{equation}
where ``$\bar{\sigma}_{*}\text{:fix}$'' in the integral represents the condition given in (\ref{eq:stress_fix-con}). 

\subsection{Saddle-point method} \label{subsec:saddle}

We now evaluate the right-hand side of (\ref{eq:pro_sigma_*}).
In the asymptotic regime $\tau\gg\tau_{\rm micro}$, the functional integral may be accurately evaluated by the saddle-point method.
By introducing the Lagrange multiplier field $\bm{\lambda}(\bm{r})=(\lambda^{r}(\bm{r}), \lambda^{\theta}(\bm{r}),\lambda^{\varphi}(\bm{r}))$ to take constraint (\ref{eq:stress_balance}) into account, we obtain
\begin{equation}
P(\bar{\sigma}_{*}) = C''' \exp \left[ -\tau I(\bar{\sigma}_{*}) \right] 
\label{eq:pro_sur_stress}
\end{equation}
with
\begin{equation}
I(\bar{\sigma}_{*}) = \min_{\substack{ \{ \hat{\sigma}^{\alpha \beta}(\bm{r})\} \\ \bar{\sigma}_{*}\text{:fix} }}\; \frac{1}{4k_{\rm B}T\eta}\int_{\mathcal{R}}^{\infty}dr \int_{0}^{\pi} d\theta \int_{0}^{2\pi} d\varphi \; \mathcal{L}\left( \{ \hat{\sigma}^{\alpha \beta}(\bm{r})\} \right) ,
  \label{eq:contraction_stress}
\end{equation}
and
\begin{equation}
 \mathcal{L}\left( \{ \hat{\sigma}^{\alpha \beta}(\bm{r})\} \right) = r^2 \sin\theta \; \Big[ \hat{s}^{{\alpha}{\beta}}(\bm{r}) (\tilde \Delta^{-1})^{{\alpha}{\beta}{\alpha}'{\beta}'} \hat{s}^{{\alpha}'{\beta}'}(\bm{r}) + 
\bm{\lambda}(\bm{r})\cdot \left( \nabla \cdot 
\overleftrightarrow{\hat{\bm{\sigma}}}(\bm{r})\right) \Big] ,
 \label{eq:Lagrangian}
\end{equation}
where $C'''$ is the normalization constant, and $\hat{s}^{{\alpha}{\beta}}(\bm{r}) =\hat{\sigma}^{{\alpha}{\beta}}(\bm{r})+\hat{p}(\bm{r}) \delta^{{\alpha}{\beta}}$ with $\hat{p}(\bm{r})=-(\hat{\sigma}^{rr}(\bm{r})+\hat{\sigma}^{\theta\theta}(\bm{r})+\hat{\sigma}^{\varphi\varphi}(\bm{r}))/3$. 
Note that the saddle-point method for the large deviation function has been rigorously verified under certain conditions. This is called the contraction principle~\cite{Touchette} in probability theory.
In \ref{sec:fluc_hydro}, we present a derivation of the probability density of $\bar{\sigma}_{*}$ within the framework of fluctuating hydrodynamics.

As a reference for the argument below, we explicitly write $\nabla \cdot \overleftrightarrow{\hat{\bm{\sigma}}}$ and $\hat{s}^{\alpha\beta} (\tilde{\Delta}^{-1})^{\alpha\beta\alpha'\beta'} \hat{s}^{\alpha' \beta'}$ as
\begin{equation}
 \nabla \cdot \overleftrightarrow{\hat{\bm{\sigma}}}=
  \begin{pmatrix}
   \frac{\rd \hat{\sigma}^{rr}}{\rd r}+\frac{1}{r}\frac{\rd \hat{\sigma}^{r\theta}}{\rd \theta} +\frac{1}{r\sin\theta}\frac{\rd \hat{\sigma}^{\varphi r}}{\rd \varphi}+ \frac{2\hat{\sigma}^{rr}-\hat{\sigma}^{\theta\theta}-\hat{\sigma}^{\varphi\varphi}}{r} + \frac{\hat{\sigma}^{r\theta}}{r\tan\theta}\\[10pt]
   \frac{\rd \hat{\sigma}^{r\theta}}{\rd r} +\frac{1}{r}\frac{\rd \hat{\sigma}^{\theta\theta}}{\rd \theta} +\frac{1}{r\sin\theta}\frac{\rd \hat{\sigma}^{\theta \varphi}}{\rd \varphi}+ \frac{\hat{\sigma}^{\theta\theta}-\hat{\sigma}^{\varphi\varphi}}{r\tan\theta}+\frac{3\hat{\sigma}^{r\theta}}{r} \\[10pt]
   \frac{\rd \hat{\sigma}^{\varphi r}}{\rd r} +\frac{1}{r}\frac{\rd \hat{\sigma}^{\theta\varphi}}{\rd \theta} +\frac{1}{r\sin\theta}\frac{\rd \hat{\sigma}^{\varphi\varphi}}{\rd \varphi}+\frac{2\hat{\sigma}^{\theta\varphi}}{r\tan\theta}+\frac{3\hat{\sigma}^{\varphi r}}{r}
 \end{pmatrix},
 \label{eq:div_stress_sphe}
\end{equation}
and 
\begin{align}
 \hat{s}^{\alpha\beta} (\tilde{\Delta}^{-1})^{\alpha\beta\alpha'\beta'} \hat{s}^{\alpha' \beta'} &=  \left( \hat{\sigma}^{rr}+\hat{p}\right)^2 + \left( \hat{\sigma}^{rr}+\hat{p}\right) \left( \hat{\sigma}^{\theta\theta}+\hat{p}\right) + \left( \hat{\sigma}^{\theta\theta}+\bar{p}\right)^2 + (\hat{\sigma}^{r\theta})^2 + (\hat{\sigma}^{\theta \varphi})^2 + (\hat{\sigma}^{\varphi r})^2
 \notag
 \\[5pt]
 &= \frac{(\hat{\sigma}^{rr})^2 +(\hat{\sigma}^{\theta\theta})^2 +(\hat{\sigma}^{\varphi\varphi})^2 -\hat{\sigma}^{rr}\hat{\sigma}^{\theta\theta}-\hat{\sigma}^{\theta\theta}\hat{\sigma}^{\varphi\varphi}-\hat{\sigma}^{\varphi\varphi}\hat{\sigma}^{rr}}{3}
 \notag
 \\
 &\qquad + (\hat{\sigma}^{r\theta})^2 + (\hat{\sigma}^{\theta \varphi})^2 + (\hat{\sigma}^{\varphi r})^2 ,
 \label{eq:stress_calc_vari}
\end{align}
respectively, where we have used (\ref{eq:decomp_stress}), (\ref{eq:pres_stress}), and (\ref{eq:Delta_inv}).

Next, we calculate the right-hand side of (\ref{eq:contraction_stress}).
We start with the variation
\begin{align}
 \delta \int_{\mathcal{R}}^{\infty}dr \int_{0}^{\pi} d\theta \int_{0}^{2\pi} d\varphi\; \mathcal{L}\left( \{ \hat{\sigma}^{\alpha \beta}(\bm{r})\} \right) &= \int_{\mathcal{R}}^{\infty}dr \int_{0}^{\pi} d\theta \int_{0}^{2\pi} d\varphi\; \left\{ \rd^{\alpha'} \left[ \frac{\rd \mathcal{L}}{\rd \left( \rd^{\alpha'}\hat{\sigma}^{\alpha \beta }\right)} \delta \hat{\sigma}^{\alpha \beta}  \right] \right.
 \notag
 \\[5pt]
 &\qquad \left.+  \left[ \frac{\rd \mathcal{L}}{\rd \hat{\sigma}^{\alpha \beta}} - \rd^{\alpha'} \left( \frac{\rd \mathcal{L}}{\rd \left( \rd^{\alpha'}\hat{\sigma}^{\alpha \beta}\right)}\right)\right] \delta \hat{\sigma}^{\alpha\beta} + \frac{\rd \mathcal{L}}{\rd \lambda^{\alpha}}\delta \lambda^{\alpha}\right\}.
 \label{eq:calc_vari}
\end{align}
The first term of the right-hand side corresponds to the surface contribution.
We find that 
\begin{equation}
 \int_{0}^{2\pi} d\varphi\; \rd^{\varphi} \left[ \frac{\rd \mathcal{L}}{\rd \left( \rd^{\varphi}\hat{\sigma}^{\alpha \beta }\right)} \delta \hat{\sigma}^{\alpha \beta}  \right] =0
\end{equation}
because of the periodic boundary of $\varphi$, and
\begin{equation}
 \int_{0}^{\pi} d\theta\; \rd^{\theta} \left[ \frac{\rd \mathcal{L}}{\rd \left( \rd^{\theta}\hat{\sigma}^{\alpha \beta }\right)} \delta \hat{\sigma}^{\alpha \beta}  \right] =0,
\end{equation}
because
\begin{equation}
 \frac{\rd \mathcal{L}}{\rd \left( \rd^{\theta}\hat{\sigma}^{\alpha \beta }\right)} = 0
\end{equation}
at $\theta=0$ and $\pi$.
Furthermore, as a result of (\ref{eq:slip_cond}), $\hat{\sigma}^{r\theta}$ and $\hat{\sigma}^{\varphi r}$ are fixed to zero at the surface.
Then, 
\begin{equation}
 \delta \hat{\sigma}^{r\theta} = \delta \hat{\sigma}^{\varphi r} = 0
\end{equation}
at $r=\mathcal{R}$.
Thus, the first term of the right-hand side of (\ref{eq:calc_vari}) becomes
\begin{equation}
 \int_{\mathcal{R}}^{\infty}dr \int_{0}^{\pi} d\theta \int_{0}^{2\pi} d\varphi\; \rd^{r} \left[ \frac{\rd \mathcal{L}}{\rd \left( \rd^{r} \hat{\sigma}^{rr }\right)} \delta \hat{\sigma}^{rr} \right].
\end{equation}
Here, we impose
\begin{equation}
 \frac{\rd \mathcal{L}}{\rd \left( \rd^{r} \hat{\sigma}^{rr }\right)} \bigg\vert_{r=\mathcal{R}}= 0
\end{equation}
as the boundary condition of the variational problem.
This is called a natural boundary condition, and explicitly gives
\begin{equation}
  \lambda^{r}(\mathcal{R},\Omega) = 0.
   \label{eq:bound_lambda}
\end{equation}
With this setup, the surface contribution of the variation vanishes, and we obtain the Euler--Lagrange equations
\begin{equation}
 \frac{\rd \mathcal{L}}{\rd \hat{\sigma}^{\alpha \beta}} - \rd^{\alpha'} \left( \frac{\rd \mathcal{L}}{\rd \left( \rd^{\alpha'}\hat{\sigma}^{\alpha \beta}\right)}\right) = 0.
  \label{eq:E-L_eq}
\end{equation}
Using (\ref{eq:Lagrangian}), (\ref{eq:div_stress_sphe}), and (\ref{eq:stress_calc_vari}), we can write (\ref{eq:E-L_eq}) as
\begin{equation}
 \begin{cases}
  \displaystyle \hat{\sigma}^{rr} = - \hat{p} + \frac{\rd \lambda^{r}}{\rd r} ,
  \\[4mm]
  \displaystyle \hat{\sigma}^{\theta \theta} = - \hat{p} + \left( \frac{1}{r}\frac{\rd \lambda^{\theta}}{\rd \theta}+\frac{\lambda^r}{r}\right) ,
  \\[4mm]
  \displaystyle \hat{\sigma}^{\varphi \varphi} = - \hat{p} + \left( \frac{1}{r\sin\theta}\frac{\rd \lambda^{\varphi}}{\rd \varphi}+\frac{\lambda^r}{r}+\frac{\lambda^{\theta}}{r\tan\theta}\right)  ,
  \\[4mm]
  \displaystyle \hat{\sigma}^{r\theta} = \frac{1}{2} \left( \frac{1}{r}\frac{\rd \lambda^r}{\rd \theta} + \frac{\rd \lambda^{\theta}}{\rd r}-\frac{\lambda^{\theta}}{r}\right) ,
  \\[4mm]
  \displaystyle \hat{\sigma}^{\theta\varphi} = \frac{1}{2} \left( \frac{1}{r\sin\theta}\frac{\rd \lambda^{\theta}}{\rd \varphi}+\frac{1}{r}\frac{\rd \lambda^{\varphi}}{\rd \theta}-\frac{\lambda^{\varphi}}{r\tan \theta}\right) ,
  \\[4mm]
  \displaystyle \hat{\sigma}^{\varphi r} = \frac{1}{2} \left( \frac{1}{r\sin\theta}\frac{\rd \lambda^{r}}{\rd \varphi}+\frac{\rd \lambda^{\varphi}}{\rd r}-\frac{\lambda^{\varphi}}{r}\right) .
 \end{cases}
 \label{eq:vari_stress}
\end{equation}
Note that (\ref{eq:pres_stress}) is equivalent to
\begin{equation}
 \nabla \cdot \bm{\lambda}=0
  \label{eq:con_incom}
\end{equation}
in the expression of (\ref{eq:vari_stress}).
By solving (\ref{eq:stress_balance}), (\ref{eq:vari_stress}), and (\ref{eq:con_incom}) with boundary conditions (\ref{eq:slip_cond}), (\ref{eq:stress_fix-con}), and (\ref{eq:bound_lambda}), we obtain the probability density of $\bar{\sigma}_{*}$.
The set of equations (\ref{eq:stress_balance}), (\ref{eq:vari_stress}), and (\ref{eq:con_incom}) coincide with the Stokes equations
\begin{align}
 \eta \nabla^2 \bm{u} &= \nabla \hat{p},
 \\
 \nabla \cdot \bm{u} &= 0,
\end{align}
when we set $\bm{\lambda}(\bm{r})=2\eta \bm{u}(\bm{r})$.
We may interpret $\bm{u}(\bm{r})$ as the macroscopic fluctuating velocity of the fluid generated by $\bar{\sigma}_{*}$.

By referring to the solution of the Stokes equations, we obtain
\begin{align}
 \lambda^{r} (\bm{r})&= \left( 1-\frac{\mathcal{R}}{r}\right) 2\mathcal{R}\bar{\sigma}_{*}\cos\theta ,
 \\
 \lambda^{\theta} (\bm{r})&= - \left( 1-\frac{\mathcal{R}}{2r}\right) 2\mathcal{R}\bar{\sigma}_{*}\sin\theta ,
 \\
 \lambda^{\varphi}(\bm{r})&= 0,
 \\
 \hat{p} (\bm{r})&= p_{\infty} - \frac{\mathcal{R}^2}{r^2}\bar{\sigma}_{*}\cos\theta ,
\end{align}
where $p_{\infty}$ is a constant.
Then, (\ref{eq:vari_stress}) leads to
\begin{align}
 \hat{\sigma}^{rr}(\bm{r}) &= -p_{\infty} + \frac{3\mathcal{R}^2 \bar{\sigma}_{*}\cos\theta}{r^2},
 \\
 \hat{\sigma}^{\theta\theta}(\bm{r}) &= \hat{\sigma}^{\varphi\varphi} (\bm{r}) = -p_{\infty} ,
 \\
 \hat{\sigma}^{r\theta}(\bm{r}) &= \hat{\sigma}^{\theta \varphi}(\bm{r})=\hat{\sigma}^{\varphi r}(\bm{r}) = 0.
\end{align}
Substituting these relations into (\ref{eq:contraction_stress}), we obtain
\begin{align}
 I(\bar{\sigma}_{*}) &= \frac{1}{4k_{\rm B}T\eta}\int_{\mathcal{R}}^{\infty}dr\; r^2 \int d\Omega \; 3\left( \frac{\mathcal{R}^2 \bar{\sigma}_{*}\cos\theta}{r^2}\right)^2
 \notag
 \\
 &= \frac{(\bar{\sigma}_{*})^2}{2}\frac{2\pi \mathcal{R}^3}{k_{\rm B}T\eta}.
\end{align}
This immediately yields
\begin{equation}
 \ave{(\bar{\sigma}_{*})^2}_{\rm eq} = \frac{k_{\rm B}T\eta}{2\pi \mathcal{R}^3 \tau}.
\end{equation}
Combining this with Kirkwood's formula (\ref{eq:kirkwood_stress}), we arrive at Stokes' law
\begin{equation}
 \gamma = \frac{(4\pi \mathcal{R}^2)^2 \tau}{2k_{\rm B}T}\ave{(\bar{\sigma}_{*})^2}_{\rm eq}=4\pi \eta \mathcal{R}.
\end{equation}

\subsection{Stokes' law for a macroscopic sphere with a rough surface}
Because we have studied a macroscopic sphere with a smooth surface represented by a spherical symmetric interaction potential, the version of Stokes' law we have obtained corresponds to that with the slip boundary condition in hydrodynamics.
Indeed, we used the boundary condition (\ref{eq:slip_cond}).
In reality, the surface of a sphere is assumed to be rough.
Although the precise microscopic description of such a surface is not simple, it should be claimed that the boundary conditions (\ref{eq:slip_cond}) are not imposed in the macroscopic fluctuation theory.
In this case, the natural boundary conditions
\begin{equation}
 \bm{\lambda}(\mathcal{R},\Omega) = \bm{0}
\end{equation}
are imposed, so that the Euler--Lagrange equation can be obtained.

By referring to the solution of the Stokes equations with these modified boundary conditions, we obtain
\begin{align}
 \lambda^{r} (\bm{r})&= \left( 1-\frac{3\mathcal{R}}{2r}+\frac{\mathcal{R}^3}{2r^3}\right) \frac{4\mathcal{R}\bar{\sigma}_{*}\cos\theta}{3},
 \\
 \lambda^{\theta}(\bm{r})&= - \left( 1-\frac{3\mathcal{R}}{4r}-\frac{\mathcal{R}^3}{4r^3}\right) \frac{4\mathcal{R}\bar{\sigma}_{*}\sin\theta}{3},
 \\
 \lambda^{\varphi} (\bm{r})&= 0,
 \\
 \hat{p} (\bm{r})&= p'_{\infty} - \frac{\mathcal{R}^2}{r^2}\bar{\sigma}_{*}\cos\theta ,
\end{align}
where $p'_{\infty}$ is a constant.
Then, (\ref{eq:vari_stress}) leads to
\begin{align}
 \hat{\sigma}^{rr}(\bm{r}) &= -p'_{\infty} + \left( \frac{3\mathcal{R}^2}{r^2}-\frac{2\mathcal{R}^4}{r^4}\right) \bar{\sigma}_{*}\cos\theta ,
 \\
 \hat{\sigma}^{\theta\theta}(\bm{r}) &= \hat{\sigma}^{\varphi\varphi}(\bm{r}) = -p'_{\infty} +\frac{\mathcal{R}^4}{r^4}\bar{\sigma}_{*}\cos\theta ,
 \\
 \hat{\sigma}^{r\theta}(\bm{r}) &= -\frac{\mathcal{R}^4}{r^4}\bar{\sigma}_{*}\sin\theta ,
 \\
 \hat{\sigma}^{\theta \varphi}(\bm{r}) &=\hat{\sigma}^{\varphi r}(\bm{r}) = 0.
\end{align}
Thus, we have that
\begin{align}
 I(\bar{\sigma}_{*}) &= \int_{\mathcal{R}}^{\infty}dr \int d\Omega \; \frac{r^2}{4k_{\rm B}T\eta} \left\{ 3\left[ \left( \frac{\mathcal{R}^2}{r^2}-\frac{\mathcal{R}^4}{r^4}\right) \bar{\sigma}_{*}\cos\theta\right]^2 + \left[ \frac{\mathcal{R}^4}{r^4}\bar{\sigma}_{*}\sin\theta\right]^2 \right\}
 \notag
 \\
 &= (\bar{\sigma}_{*})^2\frac{2\pi}{4k_{\rm B}T\eta}\int_{\mathcal{R}}^{\infty}dr\; r^2\left( \frac{2\mathcal{R}^4}{r^4}-\frac{4\mathcal{R}^6}{r^6}+\frac{10\mathcal{R}^8}{3r^8}\right)
 \notag
 \\
 &= \frac{(\bar{\sigma}_{*})^2}{2}\frac{4\pi \mathcal{R}^3}{3k_{\rm B}T\eta},
\end{align}
which leads to
\begin{equation}
 \ave{(\bar{\sigma}_{*})^2}_{\rm eq} = \frac{3k_{\rm B}T\eta}{4\pi \mathcal{R}^3 \tau} .
  \label{eq:vari_sigma_stick}
\end{equation}
The substitution of (\ref{eq:vari_sigma_stick}) into Kirkwood's formula (\ref{eq:kirkwood_stress}) yields
\begin{equation}
 \gamma = \frac{(4\pi \mathcal{R}^2)^2 \tau}{2k_{\rm B}T}\ave{(\bar{\sigma}_{*})^2}_{\rm eq}=6\pi \eta \mathcal{R}.
\end{equation}
This is Stokes' law for cases with the stick boundary condition in hydrodynamics.

\section{Concluding Remarks}

In this paper, we have derived Stokes' law from Kirkwood's formula and the Green--Kubo formula in the linear-response regime where the Stokes equations are valid.
In this derivation, we did not assume the Stokes equations to describe Stokes flow, but rather formulated the relation between the stress fluctuations in the bulk and the stress fluctuation at the surface with the aid of large deviation theory.

The heart of this derivation is the contraction principle, which is a standard technique in large deviation theory~\cite{Touchette}.
In the present case, we first used a phenomenological basis to apply the large deviation function to stress fluctuations in the bulk, and then applied the contraction principle.
See~\cite{Shiraishi} for another application of the contraction principle to the calculation of a large deviation function for time-averaged quantities.
A similar argument has been employed to derive a generating function for the current fluctuation in stochastic non-equilibrium lattice gases~\cite{Bodineau-Derrida}.
Such a phenomenological argument is called the additivity principle, and the condition of its validity can be discussed within the framework of fluctuating hydrodynamics \cite{BertiniETAL}.
Similarly, our derivation can also be formulated within fluctuating hydrodynamics, as briefly explained in \ref{sec:fluc_hydro}.
Although there is no one-to-one correspondence between our argument and that in the additivity principle, there may be a universal concept behind these two arguments.
Determining a theoretical framework that provides a microscopic understanding of the phenomenological arguments would be an interesting problem. 

In this paper, we studied the force from a viscous fluid.
More complex cases can be discussed using a similar formulation.
For example, a cross-over from the dilute case (\ref{eq:kinetic_fric_intro}) to the viscous case (\ref{eq:stokes_fric_intro}), which has been observed in numerical experiments~\cite{Lee-Kapral}, may be one of our next targets.
A simple but less trivial example may be a Brownian particle under a temperature gradient~\cite{Duhr-Braun,Piazza-Parola,JiangETAL,Wurger}.
To elucidate the mechanism of the driving force and the friction force in this system, we must consider macroscopic fluctuation theory from microscopic mechanics, where the fluctuation of the energy flux should be taken into account~\cite{FruleuxETAL}.
Furthermore, the study of the force to a small system from an active environment is a hot topic in recent non-equilibrium statistical mechanics~\cite{MarchettiETAL}. 
Examples of active environments include cytoskeleton networks (under chemical reaction)~\cite{BursacETAL,MizunoETAL,TrepatETAL,Fletcher-Mullins}, granular materials~\cite{AlbertETAL,Katsuragi-Durian,Goldman-Umbanhowar,Takehara-Okumura}, and the assembly of small biological elements~\cite{YangETAL,Takatori-Yan-Brady,GinotETAL,SolonETAL}.
The law of the force from an active environment may be given by a phenomenological description in accordance with experimental observations.
However, the nature of the force is highly non-trivial. Thus, we believe that developing the theory by which the nature of the force from an active environment can be described would be a significant achievement.

Finally, let us return to the formula described in (\ref{eq:contraction_stress}).
This claims that the large deviation function on the surface is determined from the variational principle associated with the large deviation function in the bulk.
Note that the large deviation function is called entropy, because the large deviation of the fluctuations of thermodynamic variables is equal to the thermodynamic entropy.
Here, one may recall the so-called bulk-boundary correspondence in quantum many-body systems \cite{Maldacena,Gubser-Klebanov-Polyakov,Witten,AharonyETAL,Ryu-Takayanagi}.
For the moment, we do not have any evidence for a direct connection with such theories, but it may be interesting to imagine such a possibility.

\begin{acknowledgements}
The authors would like to thank T. Matsumoto, Y. Nakayama, K. Sekimoto, and A. Yoshimori for useful discussions.
The present study was supported by KAKENHI Nos. 22340109 and 25103002, and by the JSPS Core-to-Core program ``Non-equilibrium dynamics of soft-matter and information.''
\end{acknowledgements}

\renewcommand{\thesection}{Appendix \Alph{section}}
\setcounter{section}{0}
\renewcommand{\theequation}{\Alph{section}.\arabic{equation}}
\setcounter{equation}{0}
\section{\hspace{-1.2mm}: Fluctuating hydrodynamics} \label{sec:fluc_hydro}

We describe a viscous incompressible fluid at low Reynolds number by the fluctuating hydrodynamic equations in spherical coordinates.
We denote by $\bm{u}(\bm{r},t)$, $p(\bm{r},t)$, $\overleftrightarrow{\bm{\sigma}}(\bm{r},t)$, and $\overleftrightarrow{\bm{s}}(\bm{r},t)$ the velocity, pressure, stress tensor, and random stress tensor at position $\bm{r}$ and time $t$, respectively.
In this case, each component of the stress tensor is written as
\begin{align}
 \sigma^{rr} &= - p + 2\eta \frac{\rd u^{r}}{\rd r} + s^{rr},
 \label{eq:app_stress_rr}
 \\
 \sigma^{\theta \theta} &= - p + 2\eta \left( \frac{1}{r}\frac{\rd u^{\theta}}{\rd \theta}+\frac{u^{r}}{r}\right) + s^{\theta\theta} ,
 \label{eq:app_stress_thetatheta}
 \\
 \sigma^{\varphi \varphi} &= - p + 2\eta \left( \frac{1}{r\sin\theta}\frac{\rd u^{\varphi}}{\rd \varphi}+\frac{u^{r}}{r}+\frac{u^{\theta}}{r\tan\theta}\right) + s^{\varphi\varphi} ,
 \label{eq:app_stress_phiphi}
 \\
 \sigma^{r\theta} &= \eta \left( \frac{1}{r}\frac{\rd u^{r}}{\rd \theta} + \frac{\rd u^{\theta}}{\rd r}-\frac{u^{\theta}}{r}\right) + s^{r\theta},
 \label{eq:app_stress_rtheta}
 \\
 \sigma^{\theta\varphi} &= \eta \left( \frac{1}{r\sin\theta}\frac{\rd u^{\theta}}{\rd \varphi}+\frac{1}{r}\frac{\rd u^{\varphi}}{\rd \theta}-\frac{u^{\varphi}}{r\tan \theta}\right) + s^{\theta\varphi},
 \label{eq:app_stress_thetaphi}
 \\
 \sigma^{\varphi r} &= \eta \left( \frac{1}{r\sin\theta}\frac{\rd u^{r}}{\rd \varphi}+\frac{\rd u^{\varphi}}{\rd r}-\frac{u^{\varphi}}{r}\right) + s^{\varphi r}.
 \label{eq:app_stress_phir}
\end{align}
We assume that the temperature $T$ and density $\rho$ of the fluid are constant and homogeneous.
Because we are focusing on incompressible fluid, we assume that the bulk viscosity $\zeta$ is equal to zero and that $\nabla\cdot\bm{u}=0$.
This is written as 
\begin{equation}
 \frac{\rd u^{r}}{\rd r} + \frac{2u^{r}}{r} + \frac{1}{r}\frac{\rd u^{\theta}}{\rd \theta} + \frac{u^{\theta}}{r\tan\theta} + \frac{1}{r\sin\theta}\frac{\rd u^{\varphi}}{\rd \varphi} =0.
  \label{eq:app_incomp}
\end{equation}
Furthermore, the time evolution equation of $\bm{u}(\bm{r},t)$ is assumed to be given by
\begin{equation}
 \rho \frac{\rd \bm{u}}{\rd t} = \nabla \cdot \overleftrightarrow{\bm{\sigma}},
  \label{eq:app_Stokes_eq}
\end{equation}
because the Reynolds number is sufficiently low. 
Finally, the random stresses are assumed to be zero-mean Gaussian white noises with covariance
\begin{align}
 & \ave{s^{\alpha \beta}(r,\Omega,t)s^{\alpha' \beta'}(r',\Omega',t')}
 \notag
 \\
 & \qquad = 2k_{\rm B}T\eta \Delta^{\alpha \beta \alpha' \beta'}\frac{\delta(r-r')\delta(\Omega-\Omega')}{r^2}\delta(t-t'),
  \label{eq:app_stress_noise}
\end{align}
\begin{equation}
 \Delta^{\alpha \beta \alpha' \beta'} = \delta^{\alpha \alpha'}\delta^{\beta \beta'} + \delta^{\alpha \beta'}\delta^{\beta \alpha'} - \frac{2}{3}\delta^{\alpha \beta}\delta^{\alpha' \beta'}.
\end{equation}
where we have used $\zeta = 0$.
Note that (\ref{eq:app_stress_noise}) leads to
\begin{equation}
 s^{rr}+s^{\theta\theta}+s^{\varphi\varphi}=0.
  \label{eq:app_noise_incomp}
\end{equation}
By recalling (\ref{eq:app_stress_rr}), (\ref{eq:app_stress_thetatheta}), (\ref{eq:app_stress_phiphi}), and (\ref{eq:app_incomp}), we also have
\begin{equation}
 \sigma^{rr}+\sigma^{\theta\theta}+\sigma^{\varphi\varphi} + 3p=0.
  \label{eq:app_pres_stress}
\end{equation}

For any time-dependent quantity $A(t)$, we denote its path during the time interval $[0,\tau]$ by $[A]$.
Then, using (\ref{eq:app_stress_noise}) and the Gaussian property of $s^{\alpha \beta}$, we obtain the probability density of $\{ [p],[\sigma^{\alpha \beta}],[u^{\alpha}]\}$ in the form
\begin{align}
 \mathcal{P}\left( \{ [p],[\sigma^{\alpha \beta}],[u^{\alpha}]\} \right) &= C_{0} \exp \left[ -\tau \mathcal{I}\left( \{ [p],[\sigma^{\alpha \beta}],[u^{\alpha}]\} \right) \right]
 \notag
 \\
 & \qquad \times \prod_{t}\prod_{\bm{r}}\delta \left( \rho \rd_{t} \bm{u}-\nabla\cdot\overleftrightarrow{\bm{\sigma}}\right)
 \notag
 \\
 & \qquad \times \prod_{t}\prod_{\bm{r}} \delta \left( \nabla \cdot \bm{u} \right) \delta \left( \sigma^{rr}+ \sigma^{\theta\theta}+ \sigma^{\varphi\varphi}+3p\right) 
 \label{eq:app_pro_all}
\end{align}
with
\begin{align}
 \mathcal{I} \left( \{ [p],[\sigma^{\alpha \beta}],[u^{\alpha}]\} \right) &= \frac{1}{4k_{\rm B}T\eta \tau} \int_{0}^{\tau} dt \int_{\mathcal{R}}^{\infty} dr\; r^2\int d\Omega \; s^{\alpha\beta}( \tilde{\Delta}^{-1})^{\alpha \beta \alpha' \beta'}s^{\alpha'\beta'} ,
\end{align}
where $C_0$ is the normalization constant and $s^{\alpha\beta}$ in the right-hand side is related to $(p,\sigma^{\alpha\beta},u^{\alpha})$ through (\ref{eq:app_stress_rr})--(\ref{eq:app_stress_phir}).
By applying the contraction principle to (\ref{eq:app_pro_all}), we obtain the probability density of the surface stress fluctuations as
\begin{align}
 P(\bar{\sigma}_{*}) &= \int_{\bar{\sigma}_{*}\text{:fix}} \mathcal{D}p \mathcal{D} \sigma^{\alpha \beta} \mathcal{D} u^{\alpha}\; \mathcal{P}\left( \{ [p],[\sigma^{\alpha \beta}],[u^{\alpha}]\} \right)
 \notag
 \\
 & = C_0' \exp \left[ -\tau I(\bar{\sigma}_{*}) \right]
\end{align}
with
\begin{equation}
I(\bar{\sigma}_{*}) = \min_{\substack{ \{ [\sigma^{\alpha \beta}], [u^{\alpha}]\} \\ \bar{\sigma}_{*}\text{:fix}}}\; \frac{1}{4k_{\rm B}T\eta\tau}\int_{0}^{\tau}dt \int_{\mathcal{R}}^{\infty}dr \int_{0}^{\pi} d\theta \int_{0}^{2\pi} d\varphi \; \mathcal{L}\left( \{ [\sigma^{\alpha \beta}],[u^{\alpha}]\} \right) ,
 \label{eq:app_LDF}
\end{equation}
and
\begin{align}
 \mathcal{L}\left( \{ [\sigma^{\alpha \beta}],[u^{\alpha}]\} \right) &= r^2 \sin\theta \; \Big[ s^{{\alpha}{\beta}} (\tilde \Delta^{-1})^{{\alpha}{\beta}{\alpha}'{\beta}'} s^{{\alpha}'{\beta}'} + \bm{\lambda}_{1}\cdot \left( \rho\rd_{t}\bm{u}-\nabla \cdot \overleftrightarrow{\bm{\sigma}}\right) + \lambda_2 \left( \nabla \cdot \bm{u}\right) \Big] ,
\end{align}
where $C_0'$ is the normalization constant, $\bm{\lambda}_{1}(\bm{r},t)$ and $\lambda_{2}(\bm{r},t)$ are the Lagrange multiplier fields, $p=-(\sigma^{rr}+\sigma^{\theta\theta}+\sigma^{\varphi\varphi})/3$, and ``$\bar{\sigma}_{*}\text{:fix}$'' represents the condition given in
\begin{equation}
 \bar{\sigma}_{*} = \frac{1}{4\pi \tau} \int_{0}^{\tau} dt \int d\Omega \; \left[ \cos \theta\; \sigma^{rr}(\mathcal{R},\Omega,t) - \sin \theta\; \sigma^{r\theta}(\mathcal{R},\Omega,t)\right].
\end{equation}
If a time-independent configuration with $\bm{u}=\bm{0}$ is the minimizer of (\ref{eq:app_LDF}), we obtain the same result as in (\ref{eq:pro_sur_stress})--(\ref{eq:Lagrangian}).



\begin{thebibliography}{99}
 \bibitem{Callen} Callen, H.B.: Thermodynamics and an Introduction to Thermostatistics, 2nd edn. Wiley, New York (1985)
		  
 \bibitem{Landau-LifshitzStat} Landau, L.D., Lifshitz, E.M.: Statistical Physics, Vol. 5, 3rd ed. Pergamon Press, Oxford (1980)
		 
 \bibitem{Groot-Mazur} De Groot, S.R., Mazur, P.: Non-Equilibrium Thermodynamics. Dover, New York, (1984)
 \bibitem{Zubarev} Zubarev, D.N.: Nonequilibrium Statistical Thermodynamics. Consultants Bureau, New York (1974)
 \bibitem{Kubo-Toda-Hashitsume} Kubo, R., Toda, M., Hashitsume, N.: Statistical Physics II: Nonequilibrium Statistical Mechanics. Springer-Verlag, Berlin (1985)
 \bibitem{Sasa-Tasaki} Sasa, S., Tasaki, H.: Steady state thermodynamics. J. Stat. Phys. {\bf 125}, 125--224 (2006)

 \bibitem{Landau-LifshitzFluid} Landau, L.D., Lifshitz, E.M.: Fluid Mechanics, Vol. 6, 1st ed. Pergamon Press, Oxford (1959)

 \bibitem{Chapman-Cowling} Chapman, S., Cowling, T.G.: The Mathematical Theory Of Non-Uniform Gases, 3rd ed. Cambridge University Press, Cambridge (1970)
		 
 \bibitem{Evans-Cohen-Morriss} Evans, D.J., Cohen, E.G.D., Morriss, G.P.: Probability of second law violations in shearing steady states. Phys. Rev. Lett. {\bf 71}, 2401--2404 (1993)
 \bibitem{Gallavotti-Cohen} Gallavotti, G., Cohen, E.G.D.: Dynamical ensembles in nonequilibrium statistical mechanics. Phys. Rev. Lett. {\bf 74}, 2694--2697 (1995)
 \bibitem{Kurchan} Kurchan, J.: Fluctuation theorem for stochastic dynamics. J. Phys. A {\bf 31}, 3719--3729 (1998)		 
 \bibitem{Lebowitz-Spohn} Lebowitz, J.L., Spohn, H.: A Gallavotti--Cohen-type symmetry in the large deviation functional for stochastic dynamics. J. Stat. Phys. {\bf 95}, 333--365 (1999)	 
 \bibitem{Maes} Maes, C.: The fluctuation theorem as a Gibbs property. J. Stat. Phys. {\bf 95}, 367--392 (1999)
 \bibitem{CrooksPRE1} Crooks, G.E.: Entropy production fluctuation theorem and the nonequilibrium work relation for free energy differences. Phys. Rev. E {\bf 60}, 2721--2726 (1999)
 \bibitem{CrooksPRE2} Crooks, G.E.: Path-ensemble averages in systems driven far from equilibrium. Phys. Rev. E {\bf 61}, 2361--2366 (2000)
 \bibitem{JarzynskiJSP} Jarzynski, C.: Hamiltonian derivation of a detailed fluctuation theorem. J. Stat. Phys. {\bf 98}, 77--102 (2000)

 \bibitem{JarzynskiPRL} Jarzynski, C,: Nonequilibrium equality for free energy differences. Phys. Rev. Lett. {\bf 78}, 2690--2693 (1997)


 \bibitem{Hayashi-Sasa} Hayashi, K., Sasa, S.: Linear response theory in stochastic many-body systems revisited. Physica A {\bf 370}, 407--429 (2006)
 \bibitem{SeifertRPP} Seifert, U.: Stochastic thermodynamics, fluctuation theorems and molecular machines. Rep. Prog. Phys. {\bf 75}, 126001 (2012)

 \bibitem{Sasa} Sasa, S.: Derivation of Hydrodynamics from the Hamiltonian description of particle systems. Phys. Rev. Lett. {\bf 112}, 100602 (2014)
		 
 \bibitem{Kirkwood} Kirkwood, J.: The statistical mechanical theory of transport processes I. General theory. J. Chem. Phys. {\bf 14}, 180--201 (1946)
 \bibitem{Lebowitz-Rubin} Lebowitz, J.L., Rubin, E.: Dynamical Study of Brownian Motion. Phys. Rev. {\bf 131}, 2381--2396 (1963)
 \bibitem{ZwanzigJCP} Zwanzig, R.: Elementary Derivation of Time-Correlation Formulas for Transport Coefficients. J. Chem. Phys. {\bf 40}, 2527--2533 (1964)
 \bibitem{Mazur-Oppenheim} Mazur, P., Oppenheim, I.: Molecular theory of Brownian motion. Physica {\bf 50}, 241--258 (1970)
 \bibitem{Lagarkov-Sergeev} Lagar'kov, A.N., Sergeev, V.H.: Molecular dynamics method in statistical physics. Sov. Phys. Usp. {\bf 27}, 566--588 (1978)
 \bibitem{Brey-Ordonez} Brey, J.J., Ord\'o\~nez, J.G.: Computer studies of Brownian motion in a Lennard--Jones fluid: The Stokes law. J. Chem. Phys. {\bf 76}, 3260--3263 (1982)
 \bibitem{Espanol-Zuniga} Espa\~nol, P., Z\'u\~niga, I.: Force autocorrelation function in Brownian motion theory. J. Chem. Phys. {\bf 98}, 574--580 (1993)
 \bibitem{Kaddour-Levesque} Ould-Kaddour, F., Levesque, D.: Determination of the friction coefficient of a Brownian particle by molecular-dynamics simulation. J. Chem. Phys. {\bf 118}, 7888--7891 (2003)
 \bibitem{Lee-Kapral} Lee, S.H., Kapral, R.: Friction and diffusion of a Brownian particle in a mesoscopic solvent. J. Chem. Phys. {\bf 121}, 11163--11169 (2004)
 \bibitem{Lorentz} Lorentz, H.A.: Les Theories Statistiques en Thermodynamique. B.G. Teubner, Leipzig (1912)
 \bibitem{Green1951} Green, M.S.: Brownian motion in a gas of noninteracting molecules. J. Chem. Phys. {\bf 19}, 1036--1046 (1951)
 \bibitem{Zwanzig} Zwanzig, R.: Hydrodynamic fluctuations and Stokes' law friction. J. Res. Natl. Bur. Std.(US) B {\bf 68}, 143--145 (1964)

 \bibitem{Schmitz} Schmitz, R.: Fluctuations in nonequilibrium fluids. Phys. Rep. {\bf 171}, 1--58 (1988)

 \bibitem{Kim-Karrila} Kim S., Karrila, S.J.: Microhydrodynamics: Principles and Selected Applications. Butterworth-Heinemann, Newton, MA (1991)

 \bibitem{Green1954} Green, M.S.: Markoff Random Processes and the Statistical Mechanics of Time-Dependent Phenomena. II. Irreversible Processes in Fluids. J. Chem. Phys. {\bf 22}, 398--413 (1954)

 \bibitem{Dembo-Zeitouni} Dembo, A., Zeitouni, O.: Large Deviations Techniques and Applications, 2nd edn. Springer, New York (1998)
		 
 \bibitem{Touchette} Touchette, H.: The large deviation approach to statistical mechanics. Phys. Rep. {\bf 478}, 1--69 (2009)
 \bibitem{Shiraishi} Shiraishi, N.: Anomalous System Size Dependence of Large Deviation Functions for Local Empirical Measure. J. Stat. Phys. {\bf 152}, 336--352 (2013)

 \bibitem{Bodineau-Derrida} Bodineau, T., Derrida, B.: Current fluctuations in nonequilibrium diffusive systems: an additivity principle. Phys. Rev. Lett. {\bf 92}, 180601 (2004)
 \bibitem{BertiniETAL} Bertini, L., De Sole, A., Gabrielli, D., Jona-Lasinio, G., Landim, C.: Current fluctuations in stochastic lattice gases. Phys. Rev. Lett. {\bf 94}, 030601 (2005)


 \bibitem{Duhr-Braun} Duhr, S., Braun, D.: Why molecules move along a temperature gradient. Proc. Natl. Acad. Sci. U.S.A. {\bf 103}, 19678--19682 (2006)
 \bibitem{Piazza-Parola} Piazza R., Parola, A.: Thermophoresis in colloidal suspensions. J. Phys. Condens. Matter {\bf 20}, 153102 (2008)
 \bibitem{JiangETAL} Jiang, H.R., Wada, H., Yoshinaga, N., Sano, M.: Manipulation of Colloids by a Nonequilibrium Depletion Force in a Temperature Gradient. Phys. Rev. Lett. {\bf 102}, 208301 (2009)
 \bibitem{Wurger} W{\"u}rger, A.: Thermal non-equilibrium transport in colloids. Rep. Prog. Phys. {\bf 73}, 126601 (2010)

 \bibitem{FruleuxETAL} Fruleux, A., Kawai, R., Sekimoto, K.: Momentum Transfer in Nonequilibrium Steady States. Phys. Rev. Lett. {\bf 108}, 160601 (2012)

 \bibitem{MarchettiETAL} Marchetti, M.C., Joanny, J.F., Ramaswamy, S., Liverpool, T.B., Prost, J., Rao, M., Simha, R.A.: Hydrodynamics of soft active matter. Rev. Mod. Phys. {\bf 85}, 1143--1189 (2013)

 \bibitem{BursacETAL} Bursac, P., Lenormand, G., Fabry, B., Oliver, M., Weitz, D.A., Viasnoff, V., Butlerm J.P., Fredberg, J.J.: Cytoskeletal remodelling and slow dynamics in the living cell. Nat. Mater. {\bf 4}, 557--561 (2005)
 \bibitem{MizunoETAL} Mizuno, D., Tardin, C., Schmidt, C.F., MacKintosh, F.C.: Nonequilibrium Mechanics of Active Cytoskeletal Networks. Science {\bf 315}, 370--373 (2007)
 \bibitem{TrepatETAL} Trepat, X., Deng, L., An, S.S., Navajas, D., Tschumperlin, D.J., Gerthoffer, W.T., Butler, J.P., Fredberg, J.J.: Universal physical responses to stretch in the living cell. Nature {\bf 447}, 592--595 (2007)
 \bibitem{Fletcher-Mullins} Fletcher, D.A., Mullins, R.D.: Cell mechanics and the cytoskeleton. Nature {\bf 463}, 485--492 (2010)
		 
 \bibitem{AlbertETAL} Albert, R., Pfeifer, M.A., Barab\'asi, A.-L., Schiffer, P.: Slow drag in a granular medium. Phys. Rev. Lett. {\bf 82}, 205--208 (1999)
 \bibitem{Katsuragi-Durian} Katsuragi, H., Durian, D.J.: Unified force law for granular impact cratering. Nat. Phys. {\bf 3}, 420--423 (2007)
 \bibitem{Goldman-Umbanhowar} Goldman, D.I., Umbanhowar, P.: Scaling and dynamics of sphere and disk impact into granular media. Phys. Rev. E {\bf 77}, 021308 (2008)
 \bibitem{Takehara-Okumura} Takehara, Y., Okumura, K.: High-Velocity Drag Friction in Granular Media near the Jamming Point. Phys. Rev. Lett. {\bf 112}, 148001 (2014)
		 
 \bibitem{YangETAL} Yang, X., Manning, M.L., Marchetti, M.C.: Aggregation and segregation of confined active particles. Soft matter {\bf 10}, 6477--6484 (2014)
 \bibitem{Takatori-Yan-Brady} Takatori, S.C., Yan, W., Brady, J.F.: Swim Pressure: Stress Generation in Active Matter. Phys. Rev. Lett. {\bf 113}, 028103 (2014)
 \bibitem{GinotETAL} Ginot, F., Theurkauff, I., Levis, D., Ybert, C., Bocquet, L., Berthier, L., Cottin-Bizonne, C.: Nonequilibrium Equation of State in Suspensions of Active Colloids. Phys. Rev. X {\bf 5}, 011004 (2015)
 \bibitem{SolonETAL} Solon, A.P., Fily, Y., Baskaran, A., Cates, M.E., Kafri, Y., Kardar, M., Tailleur J.: Pressure is not a state function for generic active fluids. Nature Phys. {\bf 11}, 673--678 (2015)

 \bibitem{Maldacena} Maldacena, J.M.: The large $N$ limit of superconformal field theories and supergravity. Adv. Theor. Math. Phys. {\bf 2}, 231--252 (1998)
 \bibitem{Gubser-Klebanov-Polyakov} Gubser, S.S., Klebanov, I.R., Polyakov, A.M.: Gauge theory correlators from non-critical string theory. Phys. Lett. B. {\bf 428}, 105--114 (1998)
 \bibitem{Witten} Witten, E.: Anti-de Sitter space and holography. Adv. Theor. Math. Phys. {\bf 2}, 253--291 (1998)
 \bibitem{AharonyETAL} Aharony, O., Gubser, S.S., Maldacena, J.M., Ooguri, H., Oz, Y.: Large $N$ field theories, string theory and gravity. Phys. Rep. {\bf 323}, 183--386 (2000)
 \bibitem{Ryu-Takayanagi} Ryu, S., Takayanagi, T.: Holographic Derivation of Entanglement Entropy from the anti-de Sitter Space/Conformal Field Theory Correspondence. Phys. Rev. Lett. {\bf 96}, 181602 (2006)

\end{thebibliography}
\end{document}